\documentclass[review,5p,twocolumn,authoryear]{elsarticle}

\usepackage[T1]{fontenc}
\usepackage{lmodern}
\usepackage{hyperref}
\usepackage{microtype}
\usepackage{todonotes}

\usepackage{threeparttable}
\usepackage[normalem]{ulem}

\usepackage{longtable}
\usepackage{listings}
\newcommand{\revised}[1]{#1}
\newcommand{\revisedNew}[1]{\textcolor{black}{#1}}

\usepackage{xspace}
\newcommand{\grammaroptimizer}{\textsc{GrammarOptimizer}\xspace}

\usepackage{comment}
\usepackage{url}
\usepackage{booktabs}
\renewcommand{\cite}[1]{\citep{#1}}

\usepackage{cleveref}

\usepackage[switch]{lineno}
\makeatletter
\g@addto@macro\endfrontmatter{\enlargethispage{-2\baselineskip}}
\makeatother

\begin{document}

\title{Supporting Meta-model-based Language Evolution and Rapid Prototyping with Automated Grammar Optimization}

\newif\ifanonymous
\anonymousfalse
\ifanonymous
\author{Anonymous Author(s)}
\email{anonymous@anon.com}
\affiliation{%
  \institution{Anonymous Institution}
  \city{Some city}
  \country{Some country}
}
\else
\author[1]{Weixing Zhang}
\ead{weixing.zhang@gu.se}

\author[2]{Jörg Holtmann}
\ead{joerg_holtmann@gmx.de}

\author[1,3]{Daniel Strüber}
\ead{danstru@chalmers.se}

\author[4]{Regina Hebig}
\ead{regina.hebig@uni-rostock.de}

\author[5]{Jan-Philipp Steghöfer}
\ead{jan-philipp.steghoefer@xitaso.com}

\affiliation[1]{%
  organization={Department of Computer Science \& Engineering, Chalmers | University of Gothenburg},
  city={Gothenburg},
  country={Sweden}
}
\affiliation[2]{%
  organization={Independent Researcher (main work conducted at $^a$)},
  city={Paderborn},
  country={Germany}
}
\affiliation[3]{%
  organization={Department of Software Science, Radboud University},
  city={Nijmegen},
  country={Netherlands}
}
\affiliation[4]{%
  organization={Institute of Computer Science, University of Rostock},
  city={Rostock},
  country={Germany}
}\affiliation[5]{%
  organization={XITASO GmbH IT \& Software Solutions},
  city={Augsburg},
  country={Germany}
}

\fi

\begin{abstract}
In model-driven engineering, developing a textual domain-specific language (DSL) involves constructing a meta-model, which defines an underlying abstract syntax, and a grammar, which defines the concrete syntax for the DSL.
Language workbenches such as Xtext allow the grammar to be automatically generated from the meta-model, yet
the generated grammar usually needs to be manually optimized to improve its usability. When the meta-model changes during rapid prototyping or language evolution, it can become necessary to re-generate the grammar and optimize it again, causing repeated effort and potential for errors.

In this paper, we present \grammaroptimizer, an approach for optimizing generated grammars in the context of meta-model-based language evolution. To reduce the effort for language engineers during rapid prototyping and language evolution, it offers a catalog of configurable \textit{grammar optimization rules}.
Once configured, these rules can be automatically applied and re-applied after future evolution steps, greatly reducing redundant manual effort.
In addition, some of the supported optimizations can globally change the style of concrete syntax elements, further significantly reducing the effort for manual optimizations.
The grammar optimization rules were extracted from a comparison of generated and existing, expert-created grammars, based on seven available DSLs.
An evaluation based on the seven languages shows \grammaroptimizer's ability to modify Xtext-generated grammars in a way that agrees with manual changes performed by an expert and to support language evolution in an efficient way, with only a minimal need to change existing configurations over time.
\end{abstract}

\begin{keyword}
Domain-specific Languages \sep DSL \sep Grammar \sep  Xtext \sep Language Evolution \sep Language Prototyping
\end{keyword}

\maketitle

\section{Introduction}

Domain-Specific Languages (DSLs) are a common way to describe certain application domains and to specify the relevant concepts and their relationships\revised{~\cite{van2000domain}}. 
They are, among many other things, used to describe model transformations  (the Operational transformation language of the MOF Query, View, and Transformation\,---\,QVTo~\cite{qvt} and the \mbox{ATLAS} Transformation Language\,---\,ATL~\cite{ATLSyntax}), bibliographies (BibTeX~\cite{Bibtex}), graph models (DOT~\cite{Dot}), formal requirements (the Scenario Modeling Language\,---\,SML~\cite{smlRepo} and Spectra~\cite{Spectra}), meta-models (Xcore~\cite{EclipseXcore}), or web-sites (Xenia~\cite{Xenia}).

In many cases, the syntax of the language that engineers and developers work with is textual. 
For example, DOT is based on a clearly defined and well-documented grammar so that a parser can be constructed to translate the input in the respective language into an abstract syntax tree which can then be interpreted. 

A different way to go about constructing DSLs is proposed by model-driven engineering. There, the concepts that are relevant in the domain are  captured in a meta-model which defines the \emph{abstract syntax} (see, e.g., \cite{roy2019methodology,frank2013domain,mernik2005and}). Different \emph{concrete syntaxes}, e.g., graphical, textual, or form-based, can  be defined to describe actual models that adhere to the abstract syntax. 

In this paper, we consider the Eclipse ecosystem and Xtext~\cite{bettini2016implementing} as its de-facto standard framework for developing textual DSLs.
Xtext relies on the Eclipse Modeling Framework (EMF)~\cite{steinberg2008emf} and uses its Ecore (meta-)modeling facilities as basis.
Developing a textual DSL in Xtext involves two main artifacts: a grammar, which defines the concrete syntax of the language, and a meta-model, which defines the abstract syntax.
\revised{Xtext allows either the grammar or the meta-model to be created first, and then automatically generating the one from the other (or alternatively, writing both manually and aligning them).}

\revised{Software languages change over time.
This is due to \textit{language evolution}, which entails that languages change over time to address new and changed requirements, and due to \textit{rapid prototyping}, which involves many quick iterations on an initial design.
In the case of an Xtext-based language, grammar and  meta-model  need to be modified to stay consistent with each other.
We consider two options for evolving a language in Xtext:
First, the developers can change the grammar and then use Xtext to automatically create an updated version of the meta-model from it. 
Second, the developers can change the meta-model then use Xtext to update the grammar.
We call the first approach \textit{grammar-based evolution}, and the second approach \textit{meta-model-based evolution}.}

\revised{In this paper, we focus on meta-model-based evolution, for the following rationale: While grammar-based evolution is a common way of developing languages in Xtext,
it is not geared for three scenarios that we encountered in the real world, including collaborations with an industrial partner. In particular: 
1. Several concrete syntaxes (e.g., visual, textual, tabular) for the same underlying metamodel co-exist and evolve at the same time. This is particular common in the context of blended modeling \cite{Ciccozzi.2019}, a timely modeling paradigm.
2. The metamodel comes from some external source (such as a third-party supplier or a standardization committee), which prohibits independent modification.
3. The metamodel is the central artifact of a larger ecosystem of available tools, including. e.g., automated analyses and transformations. As such, the language engineers might prefer to evolve it directly, instead of relying on the, potentially sub-optimal, output of automatically co-evolving it after grammar changes.
The real-world case that inspired this paper has aspects of the first two scenarios: we work on a language from an industry partner for which there already exists an evolving metamodel and graphical editor available.}

Compared to grammar-based evolution, meta-model-based evolution has one major disadvantage: Co-evolving the grammar after meta-model changes is more complicated than vice versa, as it involves dealing with both abstract and concrete syntax aspects, whereas updating the meta-model after grammar changes only involves abstract syntax aspects.
\revisedNew{In the state of the art, the updating needs to be done manually, which leads to effort after each evolution step.
Citing the Xtext textbook, \textit{“the drawback [of manually maintaining the Ecore model] is that you need to keep the Ecore model consistent with your DSL grammar \cite{bettini2016implementing}.}
The goal of this paper is to  substantially mitigate this disadvantage, as we will now explain.}

\revisedNew{In this paper, we propose a different approach to support meta-model-based evolution: Automated 
synchronization
of the  grammar based on simple rules, which we call \textit{grammar optimization rules}.
Such rules encode typical improvements that are made to a grammar such as the automatically created one by Xtext, e.g., changing parentheses layouts, keywords, and orders of rule fragments.
Configurations can either be automatically extracted from manual edits performed to the grammar \cite{zhang2023automated}, or explicitly created by the language engineer, as an alternative to manually  performing redundant changes affecting many places in the grammar.
Whenever the meta-model evolves, the same or a slightly modified set of optimization rules can be used to update the new grammar to have the same properties as the previous version, thus avoiding the effort for manual synchronization. }

\revised{In meta-model-based evolution scenarios, our approach can considerably reduce the manual effort for optimizations and, consequently, enable faster turnaround times.
This is due to two factors that we demonstrate in our evaluation:
First, the potential to reuse existing configurations across successive evolution steps.
For example, we considered four evolution steps from the history of QVTo.
Initially, we created a configuration that fully optimized the generated grammar to be consistent with the expert-created grammar for that evolution step. 
For the following three iterations, we only needed to modify 2, 0, and 1 configuration lines, respectively, to automatically optimize the generated grammar. Without our approach, language engineers would need to manually modify 228 lines of 66 grammar rules in each evolution step.
Second, the availability of powerful rules that enforce a large-scope change affecting many grammar rules at the same time.
 For example, for the EAST-ADL case, modifying the Xtext-generated towards the expert-created grammar required  curly braces for all attributes to be removed, while keeping the outer surrounding curly braces for each  rule. 
 Performing this change manually entails manually revising 303  rules, whereas it took only  one line of configuration in GrammarOptimizer.}

\revised{While our  approach clearly unfolds these benefits in the case of evolving languages and complex changes, it does not come for free. For locally-scoped changes, creating a configuration generally leads to more effort than a   manual  grammar edit  and hence, presents  an upfront investment that  pays off only when the language evolves over time.
In a different paper \cite{zhang2023automated}, we present an approach for automating the extraction of configurations from user-provided manual edits, thus reducing the initial manual effort to be the same as in the traditional process, while keeping the long-term benefits. Together with the present paper, for the supported kinds of changes, it supports a fully automated process for aligning the grammar  after changes to the meta-model.}

The contribution of this paper is  \grammaroptimizer, an approach that modifies a generated grammar by applying a set of configurable, modular, simple optimization rules. It integrates into the workflow of language engineers working with Eclipse, EMF, and Xtext technologies and is able to apply rules to reproduce the textual syntaxes of common, textual DSLs.

We demonstrate its applicability on seven domain-specific languages from different application areas. We also show its support for language evolution in two cases:
1), we recreate the textual model transformation language QVTo in all four versions of the official standard~\cite{qvt} with only small changes to the configuration of optimization rule applications and with high consistency of the syntax between versions; and  
2), we conceived for the automotive systems modeling language EAST-ADL~\cite{eadl} together with an industrial partner a textual concrete syntax~\cite{EATXT}, where we initially started with a grammar for a subset of the EAST-ADL meta-model (i.e., textual language version 1) and subsequently evolved the grammar to encompass the full meta-model (i.e., textual language version 2).

\revised{The remainder of this paper is structured as follows. First, in Section~\ref{sec:background}, we provide an overview of the background of this paper, in particular, on  metamodel-based textual DSL engineering. In Section~\ref{sec:related-work}, we review related research. In Section~\ref{sec:methodology}, we define the methodology of this paper. Subsequently, in Section~\ref{sec:identified-optimization-rules}, we describe the  identified optimization rules, which are the main technical contribution of this paper. Following that, in Section~\ref{sec:solution}, we present our solution of the \grammaroptimizer, which implements the identified optimization rules. In Section~\ref{sec:eval}, we present our evaluation.  Section~\ref{sec:discussion} is devoted to our discussion, where we address threats to validity, the effort required to use \grammaroptimizer, implications for practitioners and researchers, and future work. Finally, in the last section, we conclude.}

\section{Background: Textual DSL Engineering based on Meta-models}
\label{sec:background}


As outlined in the introduction, the engineering of textual DSLs can be conducted through the traditional approach of specifying grammars, but also by means of meta-models.
Both approaches have commonalities, but also differences \cite{Paige.2014}.
Like grammars specified by means of the Extended Backus Naur Form (EBNF)~\cite{iso1996iec}, meta-models enable formally specifying how the terms and structures of DSLs are composed.
In contrast to grammar specifications, however, meta-models describe DSLs as graph structures and are often used as the basis for graphical or non-textual DSLs.
Particularly, the focus in meta-model engineering is on specifying the abstract syntax. The definition of concrete syntaxes is often considered a subsequent DSL engineering step. However, the focus in grammar engineering is directly on the concrete syntax~\cite{Kleppe.2007a} and leaves the definition of the abstract syntax to the compiler.

\paragraph{Meta-model-based textual DSLs}
There are also examples of textual DSLs that are built with meta-model technology.
For example, the Object Management Group (OMG) defines textual DSLs that hook into their meta-model-based Meta Object Facility (MOF) and Unified Modeling Language ecosystems, for example, the Object Constraint Language (OCL)~\cite{OCLVersions} and the Operational transformation language of the MOF Query, View, and Transformation (QVTo)~\cite{qvt}.
However, this is done in a cumbersome way:
Both the specifications for OCL and QVTo define a meta-model specifying the abstract syntax and a grammar in EBNF specifying the concrete syntax of the DSL.
This grammar, in turn, defines a different set of concepts and, therefore, a meta-model for the concrete syntax that is different from the meta-model for the abstract syntax.
As Willink~\cite{willink2020reflections} points out, this leads to the awkward fact that the corresponding tool implementations such as Eclipse OCL \cite{EclipseOCL} and Eclipse QVTo~\cite{qvto-eclipse} also apply this distinction. 		
That is, both tool implementations each require an abstract syntax and a concrete syntax meta-model and, due to their structural divergences, a dedicated transformation between them.
Additionally, both tool implementations provide a hand-crafted concrete syntax parser, which implements the actual EBNF grammar.
Maintaining these different parts and updating the manually created ones incurs significant  effort whenever the language should be evolved.

\paragraph{Xtext.}
Xtext provides a more streamlined approach to language engineering that envisions the use of a single metamodel for defining the abstract syntax, and an associated grammar for defining the textual concrete syntax. 
Grammars are defined in a custom, EBNF-based formated.
Using an Xtext grammar, Xtext applies the ANTLR parser generator framework~\cite{antlr} to derive a parser and all its required inputs. It also generates editors along with syntax highlighting, code validation, and other useful tools.

\begin{figure}[tb]
  \centering
  \includegraphics[width=\linewidth]{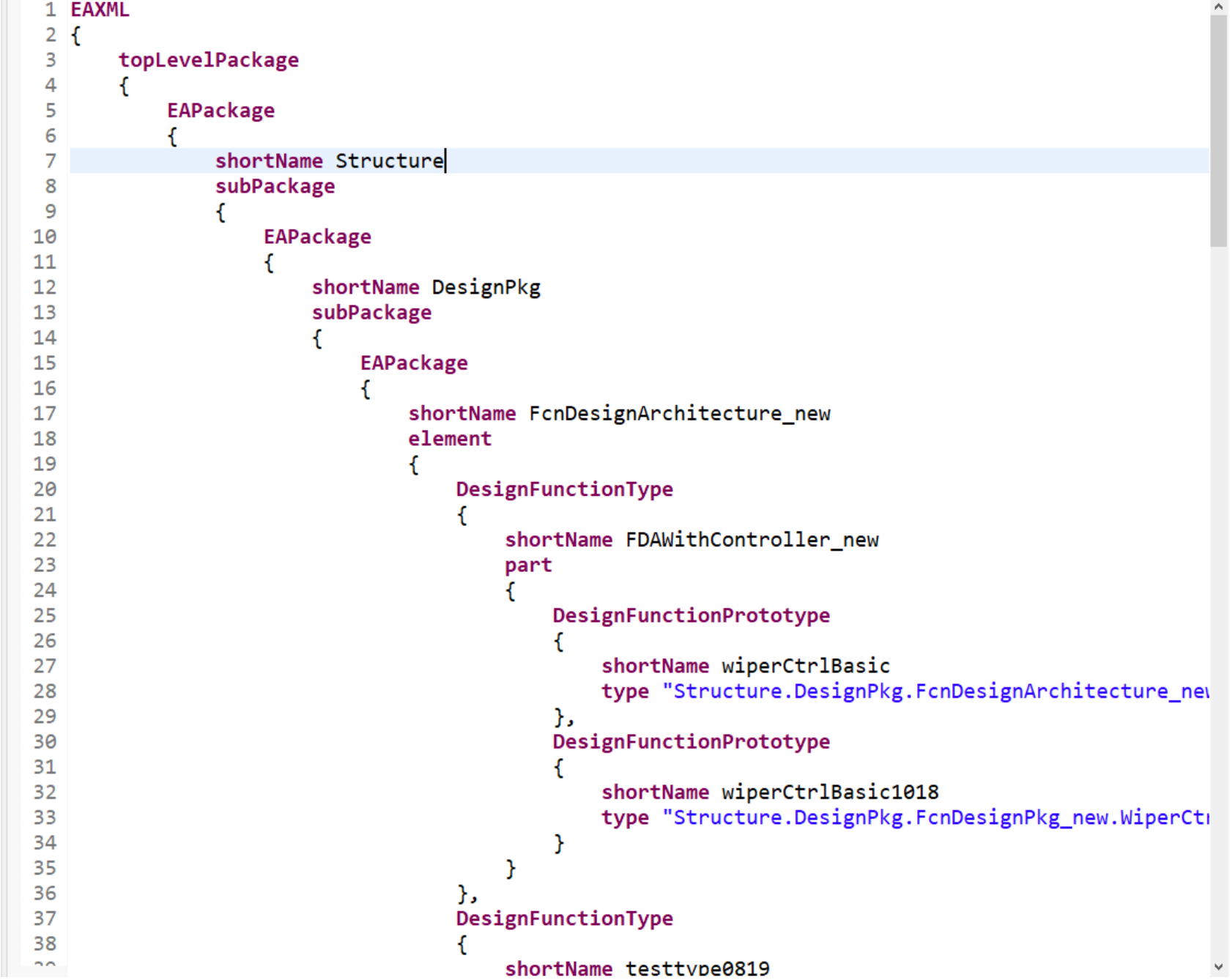}
  \caption{\revised{Instance of the generated grammar for EAST-ADL.}}
  \label{fig:example_snippet_east-adl}
\end{figure}

\begin{enumerate}
  \item \textbf{hand-craft a grammar} that maps syntactical elements of the textual concrete syntax to the concepts of the abstract syntax. this is the way many dsls have been built in xtext (e.g., xcore~\cite{eclipsexcore}, spectra~\cite{spectra}, and xenia~\cite{xenia}). however, this approach is not very robust when the meta-model changes since the grammar needs to be adapted manually to that meta-model change. 
  \item \textbf{Generate a grammar} from the meta-model using Xtext's built-in functionality (we call this grammar \emph{generated grammar} in this paper). This creates a grammar that contains grammar rules for all meta-model elements that are contained in a common root node and resolves references, etc., to a degree (see Section~\ref{sec:methodology_analysis_MMPrepsAndGrammarGen} for details). This approach deals very well with meta-model changes and only requires the re-generation of the grammar which is very fast and can be automated.
  However, the grammar is going to be very verbose, structured extensively using braces, and uses a lot of keywords. \revised{Such a situation is shown in Figure~\ref{fig:example_snippet_east-adl}, depicting an instance of the generated grammar for EAST-ADL.} Therefore, generated grammars are intended to be improved before being used in practice \cite{bettini2016implementing}.
\end{enumerate}
In this paper, we focus on making the second option more usable to give language engineers the ability to quickly re-generate their grammars when the meta-model changes, e.g., for rapid prototyping or for language evolution. Thus, we provide the ability to optimize the automatically generated grammars to improve their usability and make them similar in this regard to hand-crafted grammars. We show that this optimization can be re-applied to evolving versions of the language. 
Our contribution, \grammaroptimizer, therefore combines the advantages of both approaches while mitigating their respective disadvantages.

\section{Related Work}
\label{sec:related-work}
In the following, we discuss approaches for grammar optimization, approaches that are concerned with the design and evolution of DSLs, and other approaches.

\paragraph{Grammar Optimization}

There are a few works that aim at optimizing grammar rules with a focus on XML-based languages. For example, \citet{neubauer2015xmltext,neubauer2017xmltextToolPaper}  also mention optimization of grammar rules in Xtext. Their approach XMLText and the scope of their optimization focus only on XML-based languages. 
They convert an XML schema definition to a meta-model using the built-in capabilities of EMF.
Based on that meta-model, they then use an adapted Xtext grammar generator for XML-based languages to provide more human-friendly notations for editing XML files.
XMLText thereby acts as a sort of compiler add-on to enable editing in a different notation and to automatically translate to XML and vice versa.
In contrast, we develop a post-processing approach that enables the optimization of any Xtext grammar, not only XML-based ones,  
cf.~also our discussion in~\Cref{sec:discussion}). 

The approach of \citet{chodarev2016development} shares the same goal and a similar functional principle as XMLText, but uses other technological frameworks.
In contrast to XMLText, Chodarev supports more straightforward customization of the target XML language by directly annotating the meta-model that is generated from the XML schema.
The same distinction applies here as well: \grammaroptimizer enables the optimization of any Xtext grammar and is not restricted to XML-based languages.

Grammar optimization for DSLs in general is addressed by
\citet{Jouault.2006}. They propose an approach to specify a syntax for textual, meta-model-based DSLs with a dedicated DSL called Textual Concrete Syntax, which is based on a meta-model. 
From such a syntax specification, a concrete grammar and a parser are generated.
The approach is similar to a template language restricting the language engineer and thereby, as the authors state, lacks the freedom of grammar specifications in terms of syntax customization options. 
In contrast, we argue that the \grammaroptimizer provides more syntax customization options to achieve a well-accepted textual DSL.

Finally, \citet{novotny2013model} designed a model-driven Xtext pretty printer, which is used for improving the readability of the DSL by means of improved, language-specific, and configurable code formatting and syntax highlighting. 
In contrast, our \grammaroptimizer is not about improving code readability but focused on how to design the DSL itself to be easy to use and user-friendly.

\paragraph{Designing and Evolving Meta-model-based DSLs}
Many papers about the design of DSLs focus solely on the construction of the abstract syntax and ignore the concrete syntaxes (e.g.,~\cite{roy2019methodology,frank2011some}), or focus exclusively on graphical notations (e.g.,\cite{frank2013domain,tolvanen2018effort}).
In contrast, the guidelines proposed by \citet{karsai2014design} contain specific ideas about concrete syntax design, e.g., to ``balance compactness and comprehensibility''. 
Arguably, the languages automatically generated by Xtext are neither compact nor comprehensible and therefore require manual changes. 

\citet{mernik2005and} acknowledge that DSL design is not a sequential process. 
The paper also mentions the importance of textual concrete syntaxes to support common editing operations as well as the reuse of existing languages. 
Likewise, \citet{van2010exercise} describe DSL development as an iterative process and use EMF and Xtext for the textual syntax of the DSL. They also discuss the evolution of the language, and that ``it is hard to predict which language features will improve understandability and modifiability without actually using the language''. Again, this is an argument for the need to do prototyping when developing a language. 
\citet{karaila2009evolution} broadens the scope and also argues for the need for evolving DSLs along with the ``engineering environment'' they are situated in, including editors and code generators. 
\citet{pizka2007tool} also acknowledge the ``constant need for evolution'' of DSLs. 

There is a lot of research supporting different aspects of language change and evolution.
Existing approaches focus on how diverse artifacts can be co-evolved with evolving meta-models, namely the models that are instances of the meta-models~\cite{hebig2016approaches}, OCL constraints that are used to specify static semantics of the language~\cite{khelladi2017semi, khelladi2016metamodel}, graphical editors of the language~\cite{ruscio2010automated,di2011needed}, and model transformations that consume or produce programs of the language~\cite{garcia2012model}.
Specifically, the evolution of language instances with evolving meta-models is well supported by research approaches. For example, Di Ruscio et al.~\cite{di2011needed} support language evolution by using model transformations to simultaneously migrate the meta-model as well as model instances.

Thus, while these approaches cover a lot of requirements, there is still a need to address the evolution of textual grammars with the change of the meta-model as it happens during rapid prototyping or normal language evolution.
This is a challenge, especially since fully generated grammars are usually not suitable for use in practice. This implies that upon changing a meta-model, it is necessary to co-evolve a manually created grammar or a grammar that has been generated and then manually changed. 
\grammaroptimizer has been created to support prototyping and evolution of DSLs and is, therefore, able to support and largely automate these activities.

\paragraph{Other Approaches}
As we mentioned above, besides Xtext, there are two more approaches that support the generation of EBNF-based grammars and from these the generation of the actual parsers. These are EMFText~\cite{heidenreich2009derivation} and the Grasland toolkit~\cite{Kleppe.2007b}, which are both not maintained anymore.

Whereas our work focuses on the Eclipse technology stack based on EMF and Xtext, there are a number of other language workbenches and supporting tools that support the design of DS(M)Ls and their evolution. 
However, none of these approaches are able to derive grammars directly from meta-models, a prerequisite for the approach to language engineering we propose and the basis of our contribution, \grammaroptimizer. 
Instead, tools like textX~\cite{dejanovic2017textx} go the other way around and derive the meta-model from a grammar. Langium~\cite{langium} is the self-proclaimed Xtext successor without the strong binding to Eclipse, but does not support this particular use case just yet and instead focuses on language construction based on grammars. 
MetaEdit+~\cite{kelly2018collaborative} does not offer a textual syntax for the languages, but instead a generator to create text out of diagrams that are modeled using either tables, matrices, or diagrams. 
JetBrains MPS~\cite{mps} is based on projectional editing where concrete syntaxes are projections of the abstract syntax. 
However, these projections are manually defined and not automatically derived from the meta-model as it is the case with Xtext.
Finally, \citet{pizka2007tool} propose 
an approach to evolve DSLs including their concrete syntaxes and instances. For that, they present ``evolution languages'' that evolve the concrete syntax separately. However, they focus on DSLs that are built with classical compilers and not with meta-models. 

\section{Methodology}\label{sec:methodology}

\begin{figure*}[tb]
  \centering
  \includegraphics[width=\linewidth]{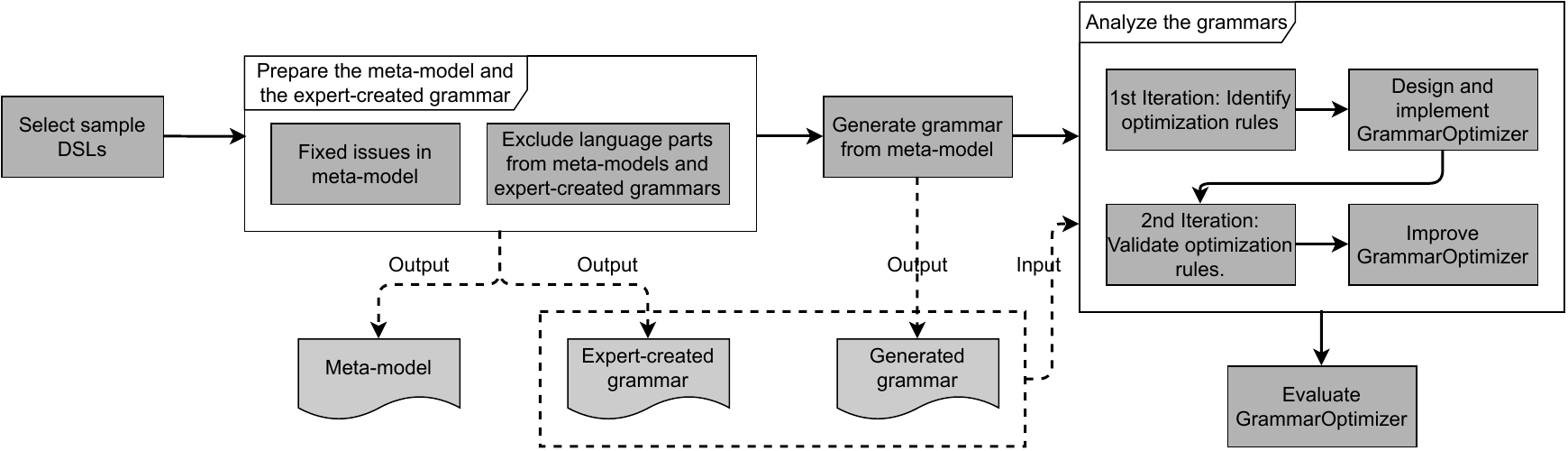}
  \caption{\revised{Overview of our methodology.}}
  \label{fig:methodology}
\end{figure*}

\revised{In this section, we describe our research methodology, shown in an overview in Fig.~\ref{fig:methodology}.
Our methodology consists of a number of sequential steps, in particular: selecting the case languages,  preparing  metamodels and grammars (including the exclusion of certain parts of the language), and  two iterations of analysis, including extraction of grammar optimization rules and tool development.
We now describe all of these steps in detail.}



\subsection{Selection of Sample DSLs}
We selected a number of DSLs for which both an expert-created grammar and a meta-model were available. Our key idea was that the expert-created grammar serves as a \textit{ground truth}, specifying what a desirable target of an optimization process would look like.
As the starting point for this optimization process, we considered the Xtext-generated grammars for the available meta-models.
The goal of our grammar optimization rules was to support an automated transformation to turn the Xtext-generated grammar into the expert-created grammar. By selecting a number of DSLs with a grammar or precise syntax definition from which we could derive such a ground truth, we aimed to generalize the grammar optimization rules so that new languages can be optimized based on rules that we include in \grammaroptimizer.

\paragraph{Sources}
To find language candidates, we collected well-known languages, such as DOT, and used language collections, such as the Atlantic Zoo \cite{Atlanticzoo}, a list of robotics DSLs~\cite{robotic2020dsls}, and similar collections~\cite{dslwikipedia,zooofdsls,dslsemanticdesign,financialdsl,van2000domain}.
However, it turned out that the search for suitable examples was not trivial despite these resources.
The quality of the meta-models in these collections was often insufficient for our purposes. In many cases, the meta-model structures were too different from the grammars or there was no grammar in either Xtext or in EBNF publicly available as well as no clear syntax definition by other means. 
We therefore extended our search to also use Github's search feature to find projects in which meta-models and Xtext grammars were present and manually searched the Eclipse Foundation's Git repositories for suitable candidates. Grammars were either taken from the language specifications or from the repositories directly.

\paragraph{Concrete Grammar Reconstruction for BibTeX}
In some cases, the syntax of a language is described in detail online, but no EBNF or Xtext grammar can be found. In our case, this is the language BibTeX. It is a well-known language to describe bibliographic data mostly used in the context of typesetting with LaTeX that is notable for its distinct syntax. In this case, we utilized the available detailed descriptions~\cite{Bibtex} to reconstruct the grammar. To validate the grammar we created, we used a number of examples of bibliographies from~\cite{Bibtex} and from our own collection to check that we covered all relevant cases.

\paragraph{Meta-model Reconstruction for DOT}
DOT is a well-known language for the specification of graph models that are input to the graph visualization and layouting tool Graphviz.
Since it is an often used language with a relatively simple, but powerful syntax, we decided to include it, even if we could not find a complete meta-model that contains both the graph structures and formatting primitives. The repository that also contains the grammar we ended up using~\cite{DotXtext}, e.g., only contains meta-models for font and graph model styles. 

Therefore, we used the Xtext grammar that parses the same language as DOT's expert-created grammar to derive a meta-model~\cite{DotXtext}.
Xtext grammars include more information than an EBNF grammar, such as information about references between concepts of the language. Thus, the fact that the DOT grammar was already formulated in Xtext allowed us to directly generate DOT's Ecore meta-model from this Xtext grammar. 
This meta-model acquisition method is an exception in this paper. Since this paper focuses on how to optimize the generated grammar, we consider this way of obtaining the meta-model acceptable for this one case.

\paragraph{Selected Cases}
As a result, we identified a sample of seven DSLs (cf.~Table~\ref{tab:analysed_DSLs}), which has a mix of different sources for meta-models and grammars. This convenience sampling consists of a mix of well-known DSLs with lesser-known, but well-developed ones. We believe this breadth of domains and language styles is broad enough to extract a generically applicable set of candidate optimization rules for \grammaroptimizer. 
%
\revised{We analyzed these selected languages in two iterations, the 1st analyzing four of them and the 2nd analyzing the remaining three.}
%
%
%
In Table~\ref{tab:analysed_DSLs}, we list all seven languages, including
information about the meta-model (source and the number of classes in the meta-model) and the expert-created grammar (source and the number of grammar rules).


\begin{table*}
\scriptsize
    \centering
    \caption{DSLs used in this paper, the sources of the meta-model and the grammar used, as well as the size of the meta-model and grammar. The first set of DSLs was analyzed to derive necessary optimization rules, and the second set to validate the candidate optimization rules and extend them if necessary.}
    \label{tab:analysed_DSLs}
    \begin{threeparttable}
    \begin{tabular}{ll l lr r r r r}
      \toprule					
       ~ & ~ 	& \multicolumn{2}{c}{\textbf{Meta-model}} & \multicolumn{2}{c}{\textbf{Expert-created Grammar}} &	\multicolumn{3}{c}{\textbf{Generated Grammar}}\\
       \cmidrule(lr){3-4}\cmidrule(lr){5-6}\cmidrule(lr){7-9}
       Iteration & DSL  	& Source 	& Classes\tnote{1} & Source & Rules & lines & {rules} & {calls}\\
        \midrule
        ~ & ATL\tnote{2} 	& Atlantic Zoo 				& 30	&  ATL Syntax 			& 28 	& 275 & 30 	& 232 \\ 
				~ & ~ 						& \cite{Atlanticzoo} 	& ~		&  \cite{ATLSyntax} & ~ 	& ~ 	& ~ 	& ~ \\
        ~ & BibTex 		& Grammarware			& 48 	& Self-built      & 46   & 293	& 48  & 188 \\ 
				1st 	& ~					& \cite{BibtexMM} & ~ 	& Based on \cite{Bibtex}								& ~ 	 & ~	& ~  & ~ \\ 
        ~ & DOT 		& Generated & 19 &   Dot          & 32 & 125	& 23  & 51 \\ 
				~ & ~ 		& ~						& ~ &   \cite{Dot}          & ~ & ~	& ~  & ~ \\ 
        ~ & SML\tnote{3} & SML repository & 48 & SML repository & 45 & 658 & 96 & 377 \\ 
				~ & ~ & \cite{smlRepo} & ~ & \cite{smlRepo} & ~ & ~ & ~ & ~ \\ 
        \midrule
        ~ & Spectra & GitHub Repository & 54 	& GitHub Repository & 58 & 442	& 62  & 243 \\
				~ & ~ 			& \cite{SpectraMM} & ~ 	& \cite{Spectra} & ~ & ~	& ~  & ~ \\
        2nd   & Xcore 	& Eclipse & 22	& Eclipse   & 26 & 243 & 33 & 149 \\ 
				~   & ~ & \cite{XcoreMM} & ~	& \cite{EclipseXcore}   & ~ & ~ & ~ & ~ \\ 
        ~ & Xenia	    & Github Repository & 13 & Github Repository  & 13 & 84 & 15  & 36 \\ 
				~ & ~	    & \cite{XeniaMM} & ~ & \cite{Xenia}  & ~ & ~ & ~  & ~ \\ 
      \bottomrule
    \end{tabular}
    \begin{tablenotes}		
        \item[1] After adaptations, containing both classes and enumerations.
        \item[2] Excluding embedded OCL rules.
        \item[3] Excluding embedded SML expressions rules.
    \end{tablenotes}
    \end{threeparttable}
    
\end{table*}

\subsection{Exclusion of Language Parts for Low-level Expressions} \label{subsubsect:Meth:exlusionParts}
Two of the analyzed languages encompass language parts for expressions, which describe low-level concepts like binary expressions (e.g., addition).
We excluded such language parts in ATL and in SML due to several aspects.
Both languages distinguish the actual language part and the expression language part already on the meta-model level and thereby treat the expression language part differently.
The respective expression parts are similarly large than the actual languages (i.e., 56 classes for the embedded OCL part of ATL and 36 classes for the SML scenario expressions meta-model), which implies a high analysis effort. 
Finally, although having a significantly large meta-model, the embedded OCL part of ATL does not specify the expressions to a sufficient level of detail (e.g., it does not allow to specify binary expressions).
\revised{Therefore, we excluded such language parts by introducing a fake class \texttt{OCLDummy}. The details for the exclusion is described in the supplemental material~\cite{datasource2023go}\footnote{See folder ``Section\_4\_Methodology''}.}


\paragraph{Exclusion from the Grammar}
In addition, we need to ensure that we can compare the language without the excluded parts to the expert-created grammar. To do so, we derive versions of the expert-created grammars in which these respective language parts are substituted by a dummy grammar rule, e.g., \texttt{OCLDummy} in the case of ATL. This dummy grammar rule is then called everywhere where a rule of the excluded language part would have been called.

\subsection{Meta-model Preparations and Generating an Xtext Grammar}\label{sec:methodology_analysis_MMPrepsAndGrammarGen}
The first step of the analysis of any of the languages is to generate an Xtext grammar based on the language's meta-model. 
This is done by using the Xtext project wizard within Eclipse.

Note that it is sometimes necessary to slightly change the meta-model to enable the generation of the Xtext grammar or to ensure that the compatibility with the expert-created grammar can be reached. 
These changes are necessary in case the meta-model is already ill-formed for EMF itself (e.g., purely descriptive Ecore files that are not intended for instantiating runtime models) or if it does not adhere to certain assumptions that Xtext makes (e.g., no bidirectional references).
\revised{The method of metamodel modification is described in detail in our  supplementary material~\cite{datasource2023go}\footnote{See directory ``Section\_4\_Methodology'.}.}

In Table~\ref{tab:analysed_DSLs}, we list how many lines, rules, and calls between rules the generated grammars included for the seven languages.

\subsection{Comparing EBNF and Xtext grammars}
\label{sec:methodology:imitation}

\revised{As a prerequisite for our analysis of grammars, we present a strategy for dealing with a noteworthy aspect of our methodology: in several cases, we dealt with languages where the expert-created grammar was available in EBNF, whereas our contribution targets Xtext, which augments EBNF with additional technicalities, such as cross-references and datatypes. 
Hence, to validate whether our approach indeed produces grammars that are equivalent to expert-created ones, 
we needed a concept that allows comparing EBNF to Xtext grammars.
}

\revised{To this end, we introduce the concept of  \emph{imitation}.
Imitation is a form of semantic equivalence of grammars that abstracts from Xtext-specific technicalities.
Specifically, we consider a set of EBNF  rules 
$\{rr_{x} | 1 \leq x \leq n\} $ to be \emph{imitated} by a set of Xtext rules  $\{ro_{y} | 1 \leq y \leq m\}$ if both produce the exact same language, modulo Xtext-specific details.
Note that the cardinalities $n$ and $m$ may differ due to situations in which one expert-created rule is replaced by several optimized rules in concert, explained below.}

\revised{Like semantic equivalence of context-free grammars, in general, \cite{hopcroft1969equivalence}, imitation is undecidable if two arbitrary grammars are considered.
However, in the scope of our analysis, we deal with specific cases that come from our evaluation subjects. These are generally of the following form:
1. Two syntactically identical---and thus, inherently semantically equivalent---grammar rules
2. Situations in which a larger rule from the first grammar is, in a controlled way, split up into several rules in the second grammar. For these, we consider them as equivalent based on a careful manual analysis, explained later.}

\begin{lstlisting}[caption=EBNF rule \texttt{edge\_stmt} from the expert-created grammar for DOT, label=DOTRealEdgeStmt,float=tb]
edge_stmt  :  (node_id | subgraph) edgeRHS [ attr_list ]
\end{lstlisting}

\begin{lstlisting}[caption=Xtext rules \texttt{EdgeStmtNode} and \texttt{EdgeStmtSubgraph} from the optimized generated grammar, label=DOTOptimizedEdgeStmt,float=tb]
EdgeStmtNode returns EdgeStmtNode:
    {EdgeStmtNode}
         node=NodeId
          (edgeRHS+=EdgeRhs)+  
          (attrLists+=AttrList)*  
    ;

EdgeStmtSubgraph returns EdgeStmtSubgraph:
    {EdgeStmtSubgraph}
         subgraph=Subgraph
          (edgeRHS+=EdgeRhs)+  
          (attrLists+=AttrList)*  
    ;
\end{lstlisting}

\revised{For example, the rule \texttt{edge\_stmt} shown in Listing~\ref{DOTRealEdgeStmt} is imitated by the combination of the rules \texttt{EdgeStmtNode} and \texttt{EdgeStmtSubgraph} shown in Listing~\ref{DOTOptimizedEdgeStmt}.
Merging the Xtext rules to form one rule, like the EBNF counterpart, was not possible in this case, due to the necessity of specifying a distinct return type in Xtext, which is not required in EBNF.
In addition, the Xtext rules contain Xtext-specific information for dealing with references and attribute types, which is not present in the EBNF rule.}

\subsection{Analysis of Grammars}
\label{sec:methodology_analysis}
We performed the analysis of existing languages in two iterations. The first iteration was purely exploratory. Here we analyzed four of the languages with the aim of finding as many candidate grammar optimization rules as possible. In the second iteration, we selected three additional languages to validate the candidate rules collected from the first iteration, add new rules if necessary, and generalise the existing rules when applicable. 

Our general approach was similar in both iterations. 
Once we had generated a grammar for a meta-model, we created a mapping between that generated grammar and the expert-created grammar of the language. The goal of this mapping was to identify which grammar rules in the generated grammar correspond to which grammar rules in the expert-created grammar.
Note that a grammar rule in the generated grammar may be mapped to multiple grammar rules in the expert-created grammar and vice versa.
From there, we inspected the generated and expert-created grammars to identify how they differed and which changes would be required to adjust the generated grammar so that it produces the same language as the expert-created grammar, i.e., \emph{imitates} the expert-created grammar rules.
We documented these changes per language and summarized them as optimization rule candidates in a spreadsheet.

For example, the expert-created grammar rule \texttt{node\_stmt} in DOT (see Listing~\ref{lst:dot-node_stmt-ebnf}) maps to the generated grammar rule \texttt{NodeStmt} in Listing~\ref{lst:dot-node_stmt-generated-xtext}.
Multiple changes are necessary to adjust the generated Xtext grammar rule:
\begin{itemize}
    \item Remove all the braces in the grammar rule \texttt{NodeStmt}.
    \item Remove all the keywords in the grammar rule \texttt{NodeStmt}.
    \item Remove the optionality from all the attributes in the grammar rule \texttt{NodeStmt}.
    \item Change the multiplicity of the attribute \texttt{attrLists} from 1..* to 0..*.
\end{itemize}

Note that in most cases the expert-created grammar was written in EBNF instead of Xtext. For example, the \texttt{returns} statement in line 1 of Listing~\ref{lst:dot-node_stmt-generated-xtext} is required for parsing in Xtext. We took that into account when comparing both grammars.

\subsubsection{First Iteration: Identify Optimization Rules}
\label{sec:methodology:analysis:first_iteration}
The analysis of the grammars of the four selected DSLs in the first iteration had two concrete purposes:
\begin{enumerate}
    \item identify the differences between the expert-created grammar and generated grammar of the language;
    \item derive grammar optimization rules that can be applied to change the generated grammar so that the optimized grammar parses the same language as the expert-created grammar.
\end{enumerate}
Please note that it is not our aim to ensure that the optimized grammar itself is identical to the expert-created grammar. 
Instead, our goal is that the optimized grammar is an \emph{imitation} of the expert-created grammar 
and therefore is able to parse the same language as the original, usually hand-crafted grammar of the DSL.
Each language was assigned to one author who performed the analysis. 

\begin{lstlisting}[caption={Non-terminal \texttt{node\_stmt} in the expert-created grammar of DOT, in \revised{EBNF}}, label={lst:dot-node_stmt-ebnf},float=tb]
node_stmt : node_id [ attr_list ]
\end{lstlisting}

\begin{lstlisting}[caption={Grammar rule \texttt{NodeStmt} in the generated grammar of DOT, in Xtext}, label={lst:dot-node_stmt-generated-xtext},float=tb]
NodeStmt returns NodeStmt:
        {NodeStmt}
        'NodeStmt'
        '{'
                ('node' node=NodeId)?
                ('attrLists' '{' attrLists+=AttrList ( "," attrLists+=AttrList)* '}' )?
        '}';
\end{lstlisting}

\begin{lstlisting}[caption={\revised{Grammar rule \texttt{NodeStmt} in the optimized grammar of DOT, in Xtext}}, label={lst:dot-node_stmt-optimized-xtext},float=tb]
NodeStmt returns NodeStmt:
    {NodeStmt}
    
    
         node=NodeId
          (attrLists+=AttrList)*  
    ;
\end{lstlisting}

As a result of the analysis, we obtained an initial set of grammar optimization rules, which contained a total of 58 candidate optimization rules. 
\Cref{tab:rules-and-iterations} summarizes in the second column the number of identified rule candidates and in the second row the number for the first iteration.
Since the initial set of grammar optimization rules was a result of an analysis done by multiple authors, it included rules that were partially overlapping and rules that turned out to only affect the grammar's formatting, but not the language specified by the grammar. Thus, we filtered rules that belong to the latter case. For rule candidates that overlapped with each other, we selected a subset of the rules as a basis for the next step. This filtering led to a selection of 46 optimization rules (cf.~third column in \Cref{tab:rules-and-iterations}).

\begin{table}
\scriptsize
\centering
\caption{Summary of identified rules their rule variants and their sources}
\label{tab:rules-and-iterations}
\begin{tabular}{@{}lp{1.5cm}p{2cm}p{1.2cm}@{}}
\toprule
\textbf{Iteration} & \textbf{Rule \newline Candidates} & \textbf{Selected Rules}  & \textbf{Rule \newline Variants} \\
\midrule
Iteration 1 & 58 & 46 & 57 \\
Iteration 2 & 10 & 10 & 10 \\
\toprule
Intermediate sum & 68 & 56 & 67\\
\bottomrule
Evaluation & 4 & 4 & 4 \\
\midrule
Overall sum & 72 & 60 & 71  \\
\bottomrule
\end{tabular}
\end{table}

We processed these 46 selected optimization rules to identify required \emph{rule variants} that could be implemented directly by means of one Java class each, which we describe more technically as part of our design and implementation elaboration in \Cref{sec:solution_ruleDesign}. 
For identifying the rule variants, we focused on the following aspects:
\begin{description}
    \item[Specification of scope] Small changes in the meta-model might lead to a different order of the lines in the generated grammar rules or even a different order of the grammar rules. Therefore, the first step was to define a suitable concept to identify the parts of the generated grammar that can function as the \emph{scope} of an optimization rule, i.e., where it applies. We identified different suitable scopes, e.g., single lines only, specific attributes, specific grammar rules, or even the whole grammar. Initially, we identified separate rule variants for each scope. Note that this also increased the number of rule variants, as for some rule candidates multiple scopes are possible.
    \item[Allowing multiple scopes] In many cases, selecting only one specific scope for a rule is too limiting. In the example above (Listing~\ref{lst:dot-node_stmt-generated-xtext}), pairs of braces in different scopes are removed: in the scope of the attribute \texttt{attrLists} in line 6 and in the scope of the containing grammar rule in lines 4 and 7. This illustrates that changes might be applied at multiple places in the grammar at once. 
    When formulating rule variants, we analyzed the rule candidates for their potential to be applied in different scopes. When suitable, we made the scope configurable. This means that only one optimization rule variant is necessary for both cases in the example. Depending on the provided parameters, it will either replace the braces for the rule or for specific attributes.
    \item[Composite optimization rules] We decided to avoid optimization rule variants that can be replaced or composed out of other rule variants, especially when such compositions were only motivated by very few cases. However, such rules might be added again later if it turns out they are needed more often. 
\end{description}

While we identified exactly one rule variant for most of the selected optimization rules, we added more than one rule variant for several of the rules. We did this when slight variations of the results were required. 
For example, we split up the optimization rule \texttt{SubstituteBrace} into the variants \texttt{ChangeBracesToParentheses}, \texttt{ChangeBracesToSquare}, and \texttt{ChangeBracesToAngle}.
Note that this split-up into variants is a design choice and not an inherent property of the optimization rule, as, e.g., the type of target bracket could be seen as nothing more than a parameter of the rule. 
As a result, we settled on 57 rule variants for the 46 identified rules (cf.~fourth column of second row in \Cref{tab:rules-and-iterations}).


\subsubsection{Second iteration: Validate Optimization Rules}
\label{sec:methodology:analysis:second_iteration}
The last step left us with 46 selected optimization rules from the first iteration (cf.~second row in \Cref{tab:rules-and-iterations}). 
We developed a preliminary implementation of \grammaroptimizer by implementing the 57 rules variants belonging to these 46 optimization rules \revised{(we will describe the implementation in the \emph{Solution} section)}. 
To validate this set of optimization rules, we performed a second iteration. In the second iteration, we selected the three DSLs Spectra, Xenia, and Xcore. As in the first iteration, we generated a grammar from the meta-model, analyzed the differences between the generated grammar and the expert-created grammar, and identified optimization rules that need to be applied to the generated grammar to accommodate these differences. 
In contrast to the first iteration, we aimed at utilizing as many existing optimization rules as possible and only added new rule candidates when necessary.

We configured the preliminary \grammaroptimizer for the new languages by specifying which optimization rules to apply on the generated grammar.
The execution results showed that the existing optimization rules were sufficient to change the generated grammar of Xenia to imitate the expert-created grammar used as the ground truth. However, we could not fully transform the generated grammar of Xcore and Spectra with the preliminary set of 46 optimization rules from the first iteration. For example, Listing~\ref{lst:xcore-xoperation-two-attributes-in-generated-grammar} shows two attributes \texttt{unordered} and \texttt{unique} in the grammar rule \texttt{XOperation} in the generated grammar for Xcore. However, in the expert-created grammar, the rule portions for the two attributes each refer to the other attribute \revised{in a way that allows using the keywords in several possible orders}, as shown  in Listing~\ref{lst:xcore-xoperation-two-attributes-in-original-grammar}. This optimization could not be performed with the optimization rules from the first iteration.

\begin{lstlisting}[caption={Two attributes in the grammar rule \texttt{XOperation} in the generated grammar of Xcore}, label={lst:xcore-xoperation-two-attributes-in-generated-grammar},float=tb]
...
        (unordered?='unordered')?
        (unique?='unique')?
...
\end{lstlisting}

\begin{lstlisting}[caption={Two attributes in the grammar rule \texttt{XOperation} in the expert-created grammar of Xcore}, label={lst:xcore-xoperation-two-attributes-in-original-grammar},float=tb]
...
          unordered?='unordered' unique?='unique'? |
          unique?='unique' unordered?='unordered'?
...
\end{lstlisting}

Based on the non-optimized parts of the grammars of Xcore and Spectra, we identified another ten optimization rules for the \grammaroptimizer. 
\revised{These ten newly identified optimization rules optimize all the non-optimized parts of the grammar of Xcore, including, e.g., optimizing the grammar in Listing~\ref{lst:xcore-xoperation-two-attributes-in-generated-grammar} to Listing~\ref{lst:xcore-xoperation-two-attributes-in-original-grammar}. These new optimization rules also optimize part of the non-optimized parts of the grammar of Spectra. We will interpret the remaining non-optimized parts in the \emph{Evaluation} section.
In the end}, after two iterations, we identified a total of 56 optimization rules (which will be implemented by a total of 67 rule variants) (cf.~fourth row in \Cref{tab:rules-and-iterations}).

\section{Identified Optimization Rules}
\label{sec:identified-optimization-rules}

In total, we identified 56 distinct optimization rules for the grammar optimization after the 2nd iteration, which we further refined into 67 rule variants (cf.~fourth row in \Cref{tab:rules-and-iterations}). 
Note that 4 additional rules were identified during the evaluation \revised{(this will be interpreted in the \emph{Evaluation} section)}, increasing the final number of identified optimization rules to 60 (cf.~bottom row in \Cref{tab:rules-and-iterations}) and the final number of rule variants to 71.

\Cref*{tab:optimizer-rules} shows some examples of the optimization rules. The rules we implemented can be categorized by the primitives they manipulate: 
grammar rules,
attributes
keywords,
braces,
multiplicities,
optionality (a special form of multiplicities),
grammar rule calls,
import statements,
symbols,
primitive types, and
lines.
They either `add' things (e.g., \emph{AddKeywordToRule}), `remove' things (e.g., \emph{RemoveOptionality}), or `change' things (e.g., \emph{ChangeCalledRule}). 
All optimization rules ensure that the resulting changed grammar is still valid and syntactically correct Xtext. 

Most optimization rules are `scoped' which means that they only apply to a specific grammar rule or attribute. In other cases, the scope is configurable, depending on the parameters of the optimization rule. For instance, the \emph{RenameKeyword} rule takes a grammar rule and an attribute as a parameter. If both are set, the scope is the given attribute in the given rule. If no attribute is set, the scope is the given grammar rule. If none of the parameters is set, the scope is the entire grammar (``Global''). All occurrences of the given keyword are then renamed inside the respective scope.

Changes to optionality are used when the generated grammar defines an element as mandatory, but the element should be optional according to the expert-created grammar. This can apply to symbols (such as commas), attributes, or keywords. Additionally, when all attributes in a grammar rule are optional, we have an optimization rule that makes the container braces and all attributes between them optional. This optimization rule allows the user of the language to enter only the grammar rule name and nothing else, e.g., ``\texttt{EAPackage DataTypes;}''.

Likewise, \grammaroptimizer contains rules to manipulate the multiplicities in the generated grammars. The meta-models and the expert-created grammars we used as inputs do not always agree about the multiplicity of elements. We provide optimization rules that can address this
within the constraints allowed by EMF and Xtext.


\begin{table}
\scriptsize
\centering
\caption{Excerpt of implemented grammar optimization rules. A configurable scope (``Config.'') means that, depending on provided parameters, the rule either applies globally to a specific grammar rule or to a specific attribute.}
\label{tab:optimizer-rules}
\begin{tabular}{@{}llll@{}}
\toprule
\textbf{Subject} & \textbf{Op.} & \textbf{Rule} & \textbf{Scope} \\
\midrule
Keyword & Add    & \emph{AddKeywordToAttr}        & Attribute \\
        &        & \emph{AddKeywordToRule}        & Rule \\
        &        & \emph{AddKeywordToLine}        & Line \\
        & Change & \emph{RenameKeyword}           & Config. \\
        &        & \emph{AddAlternativeKeyword}   & Rule \\
\midrule
Rule    & Remove & \emph{RemoveRule}              & Global \\
        & Change & \emph{RenameRule}              & Rule \\
        &        & \emph{AddSymbolToRule}         & Rule \\
\midrule
Optionality & Add & \emph{AddOptionalityToAttr}   & Attribute \\
            &     & \emph{AddOptionalityToKeyword} & Config. \\
\midrule
Import & Add    & \emph{AddImport}               & Global \\
       & Remove & \emph{RemoveImport}            & Global \\
\midrule
Brace  & Change & \emph{ChangeBracesToSquare}   & Attribute \\
       & Remove & \emph{RemoveBraces}           & Config.\\
\bottomrule
\end{tabular}
\end{table}

For the example in Listing~\ref{lst:dot-node_stmt-generated-xtext}, this means that the necessary changes to reach the same language defined in Listing~\ref{lst:dot-node_stmt-ebnf} can be implemented using the following \grammaroptimizer rules:
\begin{itemize}
    \item \emph{RemoveBraces} is applied to the grammar rule \texttt{NodeStmt} and all of its attributes. This removes all the curly braces (`\{' and `\}' in lines 4, 6, and 7) within the grammar rule.
    \item \emph{RemoveKeyword} is applied to the grammar rule \texttt{NodeStmt} and all of its attributes. This removes the keywords \texttt{`NodeStmt'}, \texttt{`node'} and \texttt{`attrLists'} (lines 3, 5, and 6) from this grammar rule.
    \item \emph{RemoveOptionality} is applied to both attributes. This removes the question marks (`?') in lines 5 and 6. 
    \item \emph{convert1toStarToStar} is applied to the attribute \texttt{attrLists}. This rule changes line 6. \revised{Before this change, this line is ``\texttt{attrLists+=AttrList ( "," attrLists+=AttrList)*}'' (the braces, keyword `\texttt{attrLists}' and the optionality `?' have been removed by previous optimization rules). After this change, it becomes \texttt{(attrLists+=AttrList)*}}.
    Note that the DOT grammar is specified using a syntax that is slightly different from standard EBNF. In that syntax, square brackets ([ and ]) enclose optional items~\cite{Dot}.
\end{itemize}
Note that line 2 in \Cref{lst:dot-node_stmt-generated-xtext} has no effect on the syntax of the grammar but is required by and specific to Xtext, so that we do not adapt such constructs. 
\revised{After the above steps, the grammar rule \texttt{NodeStmt} is adapted from Listing~\ref{lst:dot-node_stmt-generated-xtext} to Listing~\ref{lst:dot-node_stmt-optimized-xtext}.}

\section{Solution: Design and Implementation}\label{sec:solution}
The \grammaroptimizer is a Java library that offers a simple API to configure optimization rule applications and execute them on Xtext grammars. 
The language engineer can use that API to create a small program that executes \grammaroptimizer, which in turn will produce the optimized grammar.


\subsection{Grammar Representation}\label{sec:solution_grammarRepresentation}
\begin{figure}[tb]
  \centering
  \includegraphics[width=\linewidth]{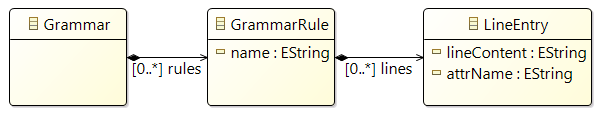}
  \caption{The class design for representing grammar rules.}
  \label{fig:class-diagram-grammar}
\end{figure}

We designed \grammaroptimizer to parse an Xtext grammar into an internal data structure which is then modified and written out again. 
%
This internal representation of the grammar follows the structure depicted in Figure~\ref{fig:class-diagram-grammar}. A \texttt{Grammar} contains a number of \texttt{GrammarRule}s that can be identified by their names. In turn, a \texttt{GrammarRule} consists of a sorted list of \texttt{LineEntry}s with their textual \texttt{lineContent} and an optional \texttt{attrName} that contains the name of the attribute defined in the line. Note that we utilize the fact that Xtext generates a new line for each attribute.

\subsection{Optimization Rule Design}\label{sec:solution_ruleDesign}
\begin{figure}[tb]
  \centering
  \includegraphics[width=\linewidth]{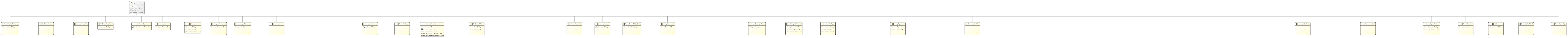}
  \caption{Excerpt of the class diagram for optimization rules.}
  \label{fig:class-diagram-optimization-rule}
\end{figure}

Internally, all optimization rules derive from the abstract class \texttt{OptimizationRule} as shown in Figure~\ref{fig:class-diagram-optimization-rule}. Derived classes overwrite the \texttt{apply()}-method to perform the specific text modifications for this rule. By doing so, the specific rule can access the necessary information through the class members: \texttt{grammar} (i.e., the entire grammar representation as explained in \Cref{sec:solution_grammarRepresentation} and depicted in \Cref{fig:class-diagram-grammar}), \texttt{grammarRuleName} (i.e., the name of the specified grammar rule that a user wants to optimize exclusively), and \texttt{attrName} (i.e., the name of an attribute that a user wants to optimize exclusively). Sub-classes can also add additional members if necessary.
This architecture makes the \grammaroptimizer extensible, as new optimization rules can easily be defined in the future. 

We built the optimization rules in a model-based manner by first creating the meta-model shown in Figure~\ref{fig:class-diagram-optimization-rule} and then using EMF to automatically generate the class bodies of the optimization rules. This way we only needed to overwrite the \texttt{apply()}-method for the concrete rules. 
Internally, the \texttt{apply()}-methods of our optimization rules are implemented using regular expressions.
Each optimization rule takes a number of parameters, e.g., the name of the grammar rule to work on or an attribute name to identify the line to work on. In addition, some optimization rules take a list of exceptions to the scope. For example, the optimization rule to remove braces can be applied to a global scope (i.e., all grammar rules) while excluding a list of specific grammar rules from the processing. This allows to configure optimization rule applications in a more efficient way.
\revised{We implemented all identified optimization rules.}\footnote{\revised{See folder `1\_Source\_Code/org.bumble.xtext.grammaroptimizer' in our supplemental material~\cite{datasource2023go}, which contains the `optimizationrule' project with the full implementation.}}
\revised{For testing, we built a comprehensive test suite, based on the optimized grammars considered in our design methodology.
We created one test case per scenario, to ensure that the grammar produced by our implementation after applying a full given configuration to an Xtext-generated grammar exactly matches an expected ground-truth grammar, for which we previously manually established that it agrees (in the sense of \textit{imitation}) with an expert-created one).}

\subsection{Configuration}

The language engineer has to configure what optimization rules the \grammaroptimizer should apply and how. This is supported by the API offered by \grammaroptimizer. 
Listing~\ref{lst:dot-config-example-rule} shows an example of how to configure the optimization rule applications in a method \texttt{executeOptimization()}, where the configuration revisits the DOT grammar optimization example transforming \Cref{lst:dot-node_stmt-generated-xtext} into \Cref{lst:dot-node_stmt-ebnf}. The lines 3 to 6 configure optimization rule applications. For example, line 3 removes all curly braces in the grammar rule \emph{NodeStmt}. 
The value of the first parameter is set to ``NodeStmt'', which means that the operation of removing curly braces will occur in the grammar rule \emph{NodeStmt}. If this first parameter is set to ``null'', the operation would be executed for all grammar rules in the grammar.
The second parameter is used to indicate the target attribute. Since it is set to ``null'', all lines in the targeted grammar rule will be affected. However, if the parameter is set to a name of an attribute, only curly braces in the line containing that attribute will be removed.
Finally, the third parameter can be used to indicate names of attributes for which the braces should not be removed. This can be used in case the second parameter is set to ``null''. 


\begin{lstlisting}[caption=Excerpt of the configuration of \grammaroptimizer{} for the QVTo 1.0 language.), label=lst:dot-config-example-rule,language=Java,float=tb]
	public static boolean executeOptimization(GrammarOptimizer go) {
   ...
   go.removeBraces("NodeStmt", null, null);
   go.removeKeyword("NodeStmt", null, null, null);
   go.removeOptionality("NodeStmt", null);
   go.convert1toStarToStar("NodeStmt", "attrLists");
   ...
	}
\end{lstlisting}

Similarly, the optimization rule application in line 4 is used to remove all keywords in the grammar rule \emph{NodeStmt}. 
Again, the second parameter can be used to specify which lines should be affected using an attribute. 
The third parameter is used to indicate the target keyword. 
Since it is set to ``null'', all keywords in the targeted lines will be removed. However, if the keyword is set, only that keyword will be removed.
The last parameter can be used to indicate names of attributes for which the keyword should not be removed. This can be used in case the second parameter is set to “null”.

Line 5 is used to remove the optionality from all lines in the grammar rule \emph{NodeStmt}.
If the second parameter gets an argument that carries the name of an attribute, the optionality is removed exclusively from the grammar line specifying the syntax for this attribute.

Finally, line 6 changes the multiplicity of the attribute \texttt{attrLists} in the grammar rule \texttt{NodeStmt} from 1..* to 0..*.
If the second parameter would get the argument ``null'', this adaptation would have been executed to all lines representing the respective attributes.

\subsection{Execution}
Once the language engineer has configured \grammaroptimizer, they can invoke the tool using \texttt{GrammarOptimizerRunner} on the command line and providing the paths to the input and output grammars there.
Alternatively, instead of invoking \grammaroptimizer via the command line and modifying \texttt{executeOptimization()}, it is also possible to use JUnit test cases to access the API and optimize grammars in known locations. This is the approach we have followed in order to \revised{generate} the results presented in this paper.

\begin{figure*}[ht]
  \centering
  \includegraphics[page=1,width=\linewidth]{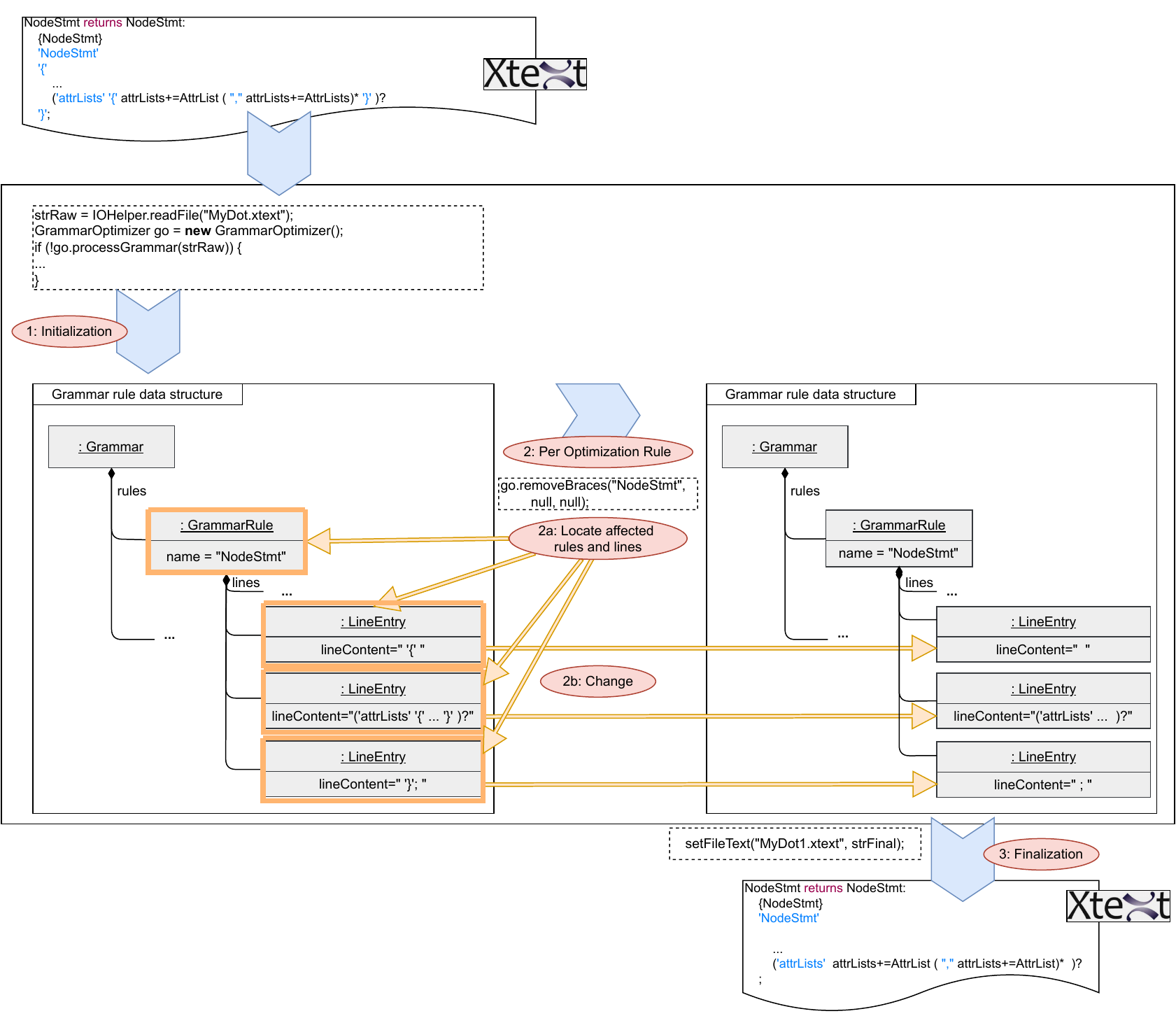}
  \caption{Exemplary Interplay of the Building Blocks of the \grammaroptimizer}
  \label{fig:GrammarOptimizer-Internal-processing-flow}
\end{figure*}

Figure~\ref{fig:GrammarOptimizer-Internal-processing-flow} uses the first optimization operation from  \Cref{lst:dot-config-example-rule} removing curly braces as an example to depict how \grammaroptimizer works internally when optimizing grammars.
The top of the figure shows an example input, which is the grammar rule \texttt{NodeStmt} generated from the meta-model of DOT (cf.~\Cref{lst:dot-node_stmt-generated-xtext}). In the lower right corner, the resulting optimized Xtext grammar rule is illustrated. \revised{In both illustrated grammar rule excerpt, blue fonts are the keywords and symbols (braces and commas).}

In \textbf{Step 1 (initialization)}, \grammaroptimizer builds a data structure out of the grammar initially generated by Xtext.
That is, it builds a \texttt{:Grammar} object containing multiple \texttt{:GrammarRule} objects, with each of them containing several \texttt{:LineEntry} objects in an ordered list.
For example, the \texttt{:Grammar} object contains a \texttt{:GrammarRule} object with the name \texttt{"NodeStmt"}.
This \texttt{:GrammarRule} object contains seven \texttt{:LineEntry} objects, which represent the seven lines of the grammar rule in \Cref{lst:dot-node_stmt-generated-xtext}. Three of these \texttt{:LineEntry} objects contain at least one curly brace (\texttt{`` `\{' ''} or \texttt{`` `\}' ''}).
\revised{These lines are explicitly represented in order to later map relevant optimization rules to them.}
Figure \ref{fig:GrammarOptimizer-Internal-processing-flow} shows an excerpt of the object structure created for the example with the three line objects for the \texttt{NodeStmt} rule.

In \textbf{Step 2 (per Optimization Rule)} each optimization rule application is processed by executing the \texttt{apply()}-method.
For our example, the optimization rule \texttt{removeBraces} is applied via the \grammaroptimizer API as configured in line 3 of \Cref{lst:dot-config-example-rule}.

In \textbf{Step 2a (localization of affected grammar rules and lines)}, 
the grammar rule and lines that need to be changed are located, based on the configuration of the optimization rule application.
In the case of our example, the grammar rule \texttt{NodeStmt} (cf.~line 1 in \Cref{lst:dot-node_stmt-generated-xtext}) is identified. 
Then, all lines of that grammar rule are identified that include a curly brace. For example, the lines represented by \texttt{:LineEntry} objects as shown in Figure \ref{fig:GrammarOptimizer-Internal-processing-flow} are identified.

In \textbf{Step 2b (change)}, the code uses regular expressions for character-level matching and searching. If it finds curly braces surrounded by single quotes (i.e., \texttt{`` `\{' ''} and \texttt{`` `\}' ''}), it removes them.

Finally, in \textbf{Step 3 (finalization)}, the \grammaroptimizer writes the complete data structure containing the optimized grammar rules to a new file by means of the call \textsf{setFileText(...)}.

\revised{After the execution of these steps, the optimized versions of the grammar is ready for use.
The typical next step is to re-generate the parser, textual editor and other artifacts for the grammar via Xtext.
We recommend that the language engineer should systematically test the resulting grammar to check whether it matches their expectations, based on the generated artifacts and a test suite with diverse  language instances.
After evolution steps, previously developed tests can act as regression tests.}





\subsection{Post-Processing vs.\ Changing Grammar Generation}
\grammaroptimizer is designed to modify grammars that Xtext generated out of meta-models.
An alternative to this post-processing approach is to directly modify the Xtext grammar generator as, e.g., in XMLText \cite{neubauer2015xmltext,neubauer2017xmltextToolPaper}.
However, we deliberately chose a post-processing approach, because the application of conventional regular expressions enables the transferability to other recent language development frameworks like Langium~\cite{langium} or textX~\cite{dejanovic2017textx}, if they support the grammar generation from a meta-model in a future point in time. 
While the optimization rules implemented in grammar optimizer are currently tailored to the structure of Xtext grammars, \grammaroptimizer does not technically depend on Xtext and the rules could easily be adapted to a different grammar language.
Furthermore, as the implementation of an Xtext grammar generator necessarily depends on many version-specific internal aspects of Xtext, the post-processing approach using regular expressions is considerably more maintainable.

\subsection{Limitations and Caveats}
Our solution has the following limitations and caveats.

First, \grammaroptimizer works on the generated grammar, which is generated from a meta-model. This means that the meta-model must contain all the concepts that the expert-created grammar has. Otherwise, the generated grammar will lack the necessary classes or attributes. This would result in the inability to imitate the expert-created grammar. 
A feasible solution would be to expand the working scope of the \grammaroptimizer, e.g., to provide a feature to detect whether all the concepts contained in the expert-created grammar corresponding elements can be found in the meta-model. 
However, we decided against implementing such a feature for now, as we see the main use case of the \grammaroptimizer not in imitating existing grammars, but in building and maintaining new DSLs.

Second, we were not able to completely imitate one of the seven languages.
In order to do so, we would have had to provide an optimization rule that would require the \grammaroptimizer user to input a multitude of parameter options.
This would have strongly increased the effort and reduced the usability to use this one optimization rule, and the rule is only required for this one language.
Thus, we argue that a manual post-adaptation is more meaningful for this one case.
However, the inherent extensibility of the \grammaroptimizer allows to add such an optimization rule if desired.
We describe the issue in a more detailed manner in \Cref{subsubsection:eval:imtatio:results}, which summarizes the evaluation results for the grammar adaptions of the seven analyzed languages.

\revised{Third, our solution is non-commutative, that is, applying the same rules with the same parametrization, but in a different order might lead to different results. For example, if \texttt{ChangeBracesToAngle} and \texttt{ChangeBracesToSquare} are successively applied to the same grammar rule, the outcome is ``last write wins'', i.e., the rule obtains square braces.  Users should be aware of this property to ensure that the achieved outcome is consistent with their intended outcome.}

\revised{Fourth, our solution does not strive to maintain backwards comparability to previous grammar versions---in general, after rule applications, instances of the previous, un-optimized grammar can no longer be parsed. 
This lack of backwards compatibility is generally desirable, as the alternative would be support for a mixing of old and new grammar elements (e.g., changed keywords and parantheses styles) in the same instance, which would generally be confusing to the user, and lead to issues with parsing and other tool support.
However, to reduce manual effort in cases where legacy grammar instances exist, automated co-evolution of grammar instances after grammar changes is generally possible and leads to  a promising future work direction (discussed in Section~\ref{sec:discussion_coevolution_instances}).
}

\section{Evaluation}\label{sec:eval} 
In this evaluation, we focus on two research questions:
\begin{itemize}
\item \emph{RQ1: Can our solution be used to adapt generated grammars so that they produce the same language as available expert-created grammars?} \\
\revised{The goal of this question is to validate the claim that our approach can automatically perform the changes that an expert would need to do manually. To this end, we consider languages for which an expert-created grammar exists, and validate the capability of our approach to re-create an equivalent grammar. }

\item \emph{RQ2: Can our solution support the co-evolution of generated grammars when the meta-model evolves? } \\
Our original motivation for the work was to enable evolution and rapid prototyping for textual languages build with a meta-model. The aim here is to evaluate whether our approach is suitable for supporting these evolution scenarios.
\end{itemize}

In the following, we address both questions. \revised{Our supplemental material~\cite{datasource2023go} contains the source code of the implementation as well as all experiments.}

\subsection{Grammar Adaptation (RQ1)}
\label{sec:eval-grammar-adaptation}
To address the first question, we evaluate the \grammaroptimizer by transforming the generated grammars of the seven DSLs, so that they parse the same syntax as the expert-created grammars. 

\subsubsection{Cases}
Our goal is to evaluate whether the \grammaroptimizer can be used to optimize the generated grammars so that their rules imitate the rules of the expert-created grammars.
%
We reused the meta-model adaptations and generated grammars from Section~\ref{sec:methodology_analysis_MMPrepsAndGrammarGen}. 
%
Furthermore, we continued working with the versions of ATL and SML in which parts of their languages were excluded as described in Section~\ref{subsubsect:Meth:exlusionParts}.

\subsubsection{Method}
For each DSL, we wrote a configuration for the final version of \grammaroptimizer which was the result of the work described in \Cref{sec:methodology,sec:identified-optimization-rules,sec:solution}. The goal was to transform the generated grammar so as to `imitate' as many grammar rules as possible from the expert-created grammar of the DSL.
Note that this was an iterative process in which we incrementally added new optimization rule applications to the \grammaroptimizer's configuration, using the expert-created grammar as a ground truth and using our notion of `imitation' (cf.~Section~\ref{sec:methodology:imitation}) as the gold standard. Essentially, we updated the \grammaroptimizer configuration and then ran the tool before analysing the optimized grammar for imitation of the original.
We repeated the process and adjusted the \grammaroptimizer configuration until the test grammar's rules `imitated' the expert-created grammar. 
Note that in the case of \textit{Spectra}, we did not reach that point. We explain this in more detail in \Cref{subsubsection:eval:imtatio:results}.
For all experiments, we used the set of 56 optimization rules that were identified after the two iterations described in Section~\ref{sec:methodology} and as summarized in Section~\ref{sec:identified-optimization-rules}.

\revised{To verify whether the optimized grammar imitates the expert-created grammar, we adopted a manual verification method, in which we systematically compared the grammar rules in the optimized grammar with the grammar rules in the expert-created grammar. An expert-created grammar is imitated by an optimized grammar if every grammar rule in it is imitated by one (or several) grammar rules from the optimized grammar.
The procedure and results of this step are documented in our supplementary materials \cite{datasource2023go}.\footnote{
See directory `2\_Supplemental\_Material/Section\_7\_Evaluation`.}}

\subsubsection{Metrics}\label{sec:eval-grammar-adaptation_metrics}
To evaluate the optimization results of the \grammaroptimizer on the case DSLs, we assessed the following metrics.

\begin{description}
    \item[$\#$\textit{GORA}] Number of \grammaroptimizer rule applications used for the configuration. 
    \item[Grammar rules] The changes in grammar rules performed by the \grammaroptimizer when adapting the generated grammar towards the expert-created grammar. We measure these changes in terms of 
		\begin{itemize}			
			\item mod: Number of modified grammar rules 
			\item add: Number of added grammar rules 
			\item del: Number of deleted grammar rules 
		\end{itemize}
    \item[Grammar lines] The changes in the lines of the grammar performed by the \grammaroptimizer when adapting the generated grammar towards the expert-created grammar. We measure these changes in terms of
		\begin{itemize}			
			\item mod: Number of modified lines 
			\item add: Number of added lines 
			\item del: Number of deleted lines 
		\end{itemize}	

    \item[Optimized grammar] Metrics about the resulting optimized grammar. We assess
		\begin{itemize}		
			\item lines: Number of overall lines 
			\item rules: Number of grammar rules 
			\item calls: Number of calls between grammar rules 
		\end{itemize}		
	\item[$\#$\textit{iGR}] Number of grammar rules in the expert-created grammar that were successfully \emph{imitated} by the optimized grammar. 

    \item[$\#$\textit{niGR}] Number of grammar rules in the expert-created grammar that were not \emph{imitated} by the optimized grammar. 
\end{description}


\subsubsection{Results} \label{subsubsection:eval:imtatio:results}

\begin{table*}
\scriptsize
    \centering
    \caption{Result of applying the GrammarOptimizer to different DSLs (RQ1)}
    \label{tab:result-of-generalizing-to-different-DSLs}
    \begin{threeparttable}
    \begin{tabular}{@{}llrrrrrrrrrrrr@{}}
        \toprule
					~&\textbf{Optimization} 	&~    & \multicolumn{3}{c}{\textbf{Grammar Rules}}	 & \multicolumn{3}{c}{\textbf{Lines in Grammar}}	& \multicolumn{3}{c}{\textbf{Optimized Grammar}}  & ~     &~	\\
					\textbf{DSL}                &\textbf{degree}   &\textbf{$\#GORA$} 	 & \textbf{mod}  & \textbf{add}                 & \textbf{del}  & \textbf{mod}	& \textbf{add}	& \textbf{del}	& \textbf{lines}	& \textbf{rules} 	& \textbf{calls}~\tnote{1}	& \textbf{$\#iGR$} 	&\textbf{$\#niGR$} \\
        \midrule
        ATL        & Complete  & 178   & 30    & 0     & 0     & 187   & 0     & 23    & 187   & 30    & 76    & 28 & 0 \\ 
        BibTeX     & Complete  & 14    & 47    & 0     & 1     & 291   & 0     & 0     & 291   & 47    & 188   & 46 & 0 \\
        DOT        & Complete  & 79    & 24    & 1     & 3     & 112   & 2     & 0     & 114   & 25    & 41    & 13 & 0 \\
        SML        & Complete  & 421   & 40    & 5     & 56    & 267   & 18    & 2     & 285   & 45    & 121   & 44 & 0 \\
        \midrule
        Spectra    & Close     & 585   & 54    & 3     & 8     & 190   & 9     & 13    & 414   & 57    & 223   & 54 & 2 \\
        Xcore      & Complete  & 307   & 20    & 7     & 14    & 179   & 35    & 10    & 214   & 27    & 100   & 25 & 0 \\
        Xenia      & Complete  & 74    & 13    & 0     & 2     & 74    & 0     & 0     & 74    & 13    & 28    & 13 & 0 \\ 
        \bottomrule
    \end{tabular}
    \begin{tablenotes}
        \item[1] The number includes the calls to dummy OCL and dummy SML expressions. 
    \end{tablenotes}
    \end{threeparttable}
\end{table*}

Table \ref{tab:result-of-generalizing-to-different-DSLs} shows the results of applying the \grammaroptimizer to the seven DSLs. See~\Cref{tab:analysed_DSLs} for the corresponding metrics of the initially generated grammars.

\paragraph{Imitation}
For all case DSLs in the first two iterations except \textit{Spectra}, we were able to achieve a complete adaptation, i.e., we were able to modify the grammar by using \grammaroptimizer so that the grammar rules of the optimized grammar \emph{imitate} all grammar rules of the expert-created grammar. 

\paragraph{Limitation regarding Spectra}
\begin{lstlisting}[caption={Example\,---\,grammar rule \texttt{TemporalPrimaryExpr} in the generated grammar of Spectra}, label={lst:excerpt-of-generated-grammar-in-Spectra},float=tb]
TemporalPrimaryExpr returns TemporalPrimaryExpr:
	{TemporalPrimaryExpr}
	'TemporalPrimaryExpr'
	'{'
		('operator' operator=EString)?
		('predPatt' predPatt=[PredicateOrPatternReferrable|EString])?
		('pointer' pointer=[Referrable|EString])?
		('regexpPointer' regexpPointer=[DefineRegExpDecl|EString])?
		('predPattParams' '{' predPattParams+=TemporalExpression ( "," predPattParams+=TemporalExpression)* '}' )?
		('tpe' tpe=TemporalExpression)?
		('index' '{' index+=TemporalExpression ( "," index+=TemporalExpression)* '}' )?
		('temporalExpression' temporalExpression=TemporalExpression)?
		('regexp' regexp=RegExp)?
	'}';
\end{lstlisting}

\begin{lstlisting}[caption={Example\,---\,grammar rule \texttt{TemporalPrimaryExpr} in the expert-created grammar of Spectra}, label={lst:excerpt-of-original-grammar-in-Spectra},float=tb]
TemporalPrimaryExpr returns TemporalExpression:
	Constant | '(' QuantifierExpr ')' | {TemporalPrimaryExpr}
	(predPatt=[PredicateOrPatternReferrable] 
	 ('(' predPattParams+=TemporalInExpr (',' predPattParams+=TemporalInExpr)* ')' | '()') | 
	 operator=('-'|'!') tpe=TemporalPrimaryExpr | 
	 pointer=[Referrable]('[' index+=TemporalInExpr ']')* | 
	 operator='next' '(' temporalExpression=TemporalInExpr ')' | 
	 operator='regexp' '(' (regexp=RegExp | regexpPointer=[DefineRegExpDecl]) ')' | 
	 pointer=[Referrable] operator='.all' | 
	 pointer=[Referrable] operator='.any' | 
	 pointer=[Referrable] operator='.prod' | 
	 pointer=[Referrable] operator='.sum' | 
	 pointer=[Referrable] operator='.min' | 
	 pointer=[Referrable] operator='.max');
\end{lstlisting}

For one of the languages, Spectra, we were able to come very close to the expert-created grammar. 
Many grammar rules of Spectra could be nearly imitated. However, we did not implement all grammar rules that would have been necessary to allow the full optimization of Spectra.
Listing~\ref{lst:excerpt-of-generated-grammar-in-Spectra} shows the grammar rule \texttt{TemporalPrimaryExpr} in Spectra's generated grammar, while Listing~\ref{lst:excerpt-of-original-grammar-in-Spectra} shows what that grammar rule looks like in the expert-created grammar. In order to optimize the grammar rule \texttt{TemporalPrimaryExpr} from Listing~\ref{lst:excerpt-of-generated-grammar-in-Spectra} to Listing~\ref{lst:excerpt-of-original-grammar-in-Spectra}, we need to configure the \grammaroptimizer so that it combines the attribute \texttt{pointer} and \texttt{operator} multiple times, and the default value of the attribute \texttt{operator} is different each time. The language engineers using the \grammaroptimizer need to input multiple parameters to ensure that the \grammaroptimizer gets enough information, and this complex optimization requirement only appears in Spectra. Therefore we did not do such an optimization.

\paragraph{Size of the Changes}
It is worth noting that the number of optimization rule applications is significantly larger than the number of grammar rules for all cases but BibTeX. This indicates that the effort required to describe the optimizations once is significant. 
However, the actual changes to the grammar, e.g., in terms of modified lines in the grammar are in most cases comparable to the number of optimization rule applications (e.g., for ATL with 178 optimization rule applications and 187 changed lines in the grammar) or even much larger (e.g., for BibTeX with 14 optimization rule applications and 291 modified lines).
Note that the number of changed, added, and deleted lines is also an underestimation of the amount of necessary changes, as many lines will be changed in multiple ways, e.g., by changing keywords and braces in the same line. This explains why for some languages the number of optimization rule applications is bigger than the number of changed lines (e.g., for SML we specified 421 optimization rule applications which changed, added, and deleted together 287 lines in the grammar).

\paragraph{Effort for the Language Engineer}
We acknowledge that the number of optimization rule applications that are necessary to adapt a generated grammar to imitate the expert-created grammar indicates that it is more effort to configure \grammaroptimizer than to apply the desired change in the grammar manually once. However, even with that assumption, we argue that the effort of configuring \grammaroptimizer is in the same order of magnitude as the effort of applying the changes manually to the grammar.

Furthermore, we argue that it is more efficient to configure \grammaroptimizer once than to manually rewrite grammar rules every time the language changes -- under the assumption that the configuration can be reused for new versions of the grammar.
In that case, the effort invested in configuring \grammaroptimizer would quickly pay off when a language is going through changes, e.g., while rapidly prototyping modifications or when the language is evolving. 
In the next section (Section \ref{sec:support_eval}), we evaluate this assumption.


In terms of reusability of the configurable optimization rules, we observe that most of the languages we cover require at least one \emph{unique} optimization rule that is not needed by any other language. This applies to DOT, BibTeX, ATL with one unique optimization rule, each. Spectra was our most complicated case with six unique rules, whereas Xcore requires four and SML requires five unique rules. This indicates that using \grammaroptimizer for a new language might require effort by implementing a few new optimization rules. However, we argue that this effort will be reduced as more optimization rules are added to \grammaroptimizer and that, in particular for evolving languages, the small investment to create a new optimization rule will pay off quickly.

\subsection{Supporting Evolution (RQ2)}\label{sec:support_eval}
To address the second question, we evaluate the \grammaroptimizer on two languages' evolution histories: The industrial case of EAST-ADL and the evolution of the DSL QVTo. We focus on the question to what degree a configuration of the \grammaroptimizer that was made for one language version can be applied to a new version of the language.

\subsubsection{Cases}
The two cases we are using to evaluate how \grammaroptimizer supports the evolution of a DSL are a textual variant of EAST-ADL~\cite{eadl} and QVT Operational (QVTo)~\cite{qvt}.

\paragraph{EAST-ADL}
EAST-ADL is an architecture description language used in the automotive domain~\cite{eadl}. 
Together with an industrial language engineer for EAST-ADL, we are currently developing a textual notation for version 2.2 of the language~\cite{EATXT}. We started this work with a simplified version of the meta-model to limit the complexity of the resulting grammar. In a later step, we switched to the full meta-model. We treat this switch as an evolution step here. The meta-model of EAST-ADL is taken from the EATOP repository~\cite{eatop-bitbucket}. The meta-model of the simplified version contains 91 classes and enumerations, and the meta-model of the full version contains 291 classes and enumerations.

\paragraph{QVTo}
QVTo is one of the languages in the OMG QVT standard~\cite{qvt}. We use the original meta-models available in Ecore format on the OMG website~\cite{qvt}. The baseline version is QVTo 1.0~\cite{qvt10spec} and we simulate evolution to version 1.1~\cite{qvt11spec}, 1.2~\cite{qvt12spec} and 1.3~\cite{qvt13spec}.
Our original intention was to use the Eclipse reference implementation of QVTo~\cite{qvto-eclipse}, but due to the differences in abstract syntax and concrete syntax (see~\Cref{sec:background}), 
we chose to use the official meta-models instead. 
We analyzed four versions of QVTo's OMG official Ecore meta-model. There are 50 differences between the meta-models of version 1.0 and 1.1, 29 of which are parts that do not contain OCL (as for ATL as described in \Cref{subsubsect:Meth:exlusionParts}, we exclude OCL in our solution for QVTo). These 29 differences include different types, for example, 1) the same set of attributes has different arrangement orders in the same class in different versions of the meta-model; 2) the same class has different superclasses in different versions; 3) the same attribute has different multiplicities in different versions, etc. There are 3 differences between versions 1.1 and 1.2, all of which are from the OCL part. There is only one difference between versions 1.2 and 1.3, and it is about the same attribute having a different lower bound for the multiplicity in the same class in the two versions.
Altogether we observed 54 meta-model differences in QVTo between the different versions \revised{(cf. the file ``Comparison of QVTo metamodel versions'' in the folder ``Section\_7\_Evaluation/Subsection\_7.2\_Support'' lists all the metamodel differences)}.

The OMG website provides an EBNF grammar for each version of QVTo, which is the basis for our imitations of the QVTo languages. Among them, versions 1.0, 1.1, and 1.2 share the same EBNF grammar for the QVTo part except for the OCL parts, despite the differences in the meta-model. The EBNF grammar of QVTo in version 1.3 is different from the other three versions.


\subsubsection{Preparation of the QVTo Case}\label{sec:support_eval_QVTo_Preparation}
In contrast to the EAST-ADL case, we needed to perform some preparations of the grammar and the meta-model to study the QVTo case. All adaptations were done the same way on all versions of QVTo.

\paragraph{Exclusion of OCL}
As described in detail in \Cref{subsubsect:Meth:exlusionParts}, we excluded the embedded OCL language part from QVTo. 
For the meta-model, we introduced a dummy class for OCL, changed all calls to OCL types into calls to that dummy class, and removed the OCL metaclasses from the meta-model.

As described in Section~\ref{subsubsect:Meth:exlusionParts}, excluding a language part such as the embedded OCL from the scope of the investigation also implies that we need to exclude this language part when it comes to judging whether a grammar is imitated. 
Therefore, we substituted all grammar rules from the excluded OCL part with a placeholder grammar rule called \texttt{ExpressionGO} where an OCL grammar rule would have been called.
This change allows us to compare the expert-created grammar of the different QVTo versions to the optimized grammar versions.

\paragraph{QVTo Meta-model Adaptations}
We found that some non-terminals of QVTo's EBNF grammar are missing in the QVTo meta-model provided by OMG. For example, there is a non-terminal \texttt{<top\_level>} in the EBNF grammar, but there is no counterpart for it in the meta-model. Therefore, we need to adapt the meta-model to ensure that it contains all the non-terminals in the EBNF grammar. 
To ensure that the adaptation of the meta-model is done systematically, we defined seven general adaptation rules that we followed when adapting the meta-models of the different versions. We list these adaptation rules in the supplemental material~\cite{datasource2023go}. 

As a result, we added 62 classes and enumerations with their corresponding references to each version of the meta-model. Note that this number is high compared to the original number of classes in the meta-model (24 classes). 
This massive change was necessary, because the available Ecore meta-models were too abstract to cover all elements of the language. The original meta-model did contain most key concepts, but would not allow to actually specify a complete QVTo transformation. For example, with the original meta-model, it was not possible to represent the scope of a mapping or helper. 

These changes enable us to imitate the QVTo grammar. 
However, they do not bias the results concerning the effects of the observed meta-model evolution as, with exception of a single case, these evolutionary differences are neither erased nor increased by the changes we performed to the meta-model. The exception is a meta-model evolution change between version 1.0 and 1.1 where the class \texttt{MappingOperation} has super types \texttt{Operation} and \texttt{NamedElement}, while the same class in V1.1 does not. 
The meta-model change performed by us removes the superclass \texttt{Operation} from \texttt{MappingOperation} in version 1.0. We did this change to prevent conflicts as the attribute \emph{name} would have been inherited multiple times by \texttt{MappingOperation}. This in turn would cause problems in the generation process. Thus, only two of the 54 meta-model evolutionary differences could not be studied. The differences and their analysis can be found in the supplemental material~\cite{datasource2023go}.



\subsubsection{Method}
To evaluate how \grammaroptimizer supports the evolution of meta-models we look at the effort that is required to update the optimization rule applications after an update of the meta-models of EAST-ADL and QVTo. 

\paragraph{Baseline \grammaroptimizer Configuration}
First, we generated the grammar for the initial version of a language's meta-model (i.e., the simple version for EAST-ADL and version 1.0 for QVTo). Then we defined the configuration of optimization rule applications that allows the \grammaroptimizer to modify the generated grammar so that its grammar rules \emph{imitate} the expert-created grammar for each case. 
Doing so confirmed the observation from the first part of the evaluation that a new language of sufficient complexity requires at least some new optimization rules (see Section \ref{subsubsection:eval:imtatio:results}). Consequently, we identified the need for four additional optimization rules for QVTo, which we implemented accordingly as part of the \grammaroptimizer (this is also summarized in Section~\ref{sec:identified-optimization-rules} in \Cref{tab:rules-and-iterations}).  
This step provided us with a baseline configuration for the \grammaroptimizer.

\paragraph{Evolution}
For the following language versions, i.e., the full version of EAST-ADL and QVTo 1.1, we then generated the grammar from the corresponding version of the meta-model and applied the \grammaroptimizer with the configuration of the previous version (i.e., simple EAST-ADL and QVTo 1.0). 
We then identified whether this was already sufficient to \emph{imitate} the language's grammar or whether changes and additions to the optimization rule applications were required. We continued adjusting the optimization rule applications accordingly to gain a \grammaroptimizer configuration valid for the new version (full EAST-ADL and QVTo 1.1, respectively). 
For QVTo, we repeated that process two more times: For QVTo 1.2, we took the configuration of QVTo 1.1 as a baseline, and for QVTo 1.3, we took the configuration of QVTo 1.2 as a baseline.


\subsubsection{Metrics}
We documented the metrics used in Section~\ref{sec:eval-grammar-adaptation_metrics} for EAST-ADL and QVTo in their different versions. In addition, we also documented the following metric:
\begin{description}
    \item [$\#$\textit{cORA}] The number of changed, added, and deleted optimization rule applications compared to the previous language version.
\end{description}

\subsubsection{Results}
Table~\ref{tab:result-of-supporting-evolution} shows the results of the evolution cases. 

\paragraph{EAST-ADL}
Compared with the simplified version of EAST-ADL, the full version is much larger. It contains 291 metaclasses, i.e., 200 metaclasses more than the simple version of EAST-ADL, which leads to a generated grammar with 291 grammar rules and 2,839 non-blank lines in the generated grammar file (cf.~\Cref{tab:result-of-supporting-evolution}).

The 22 optimization rule applications for the simple version of EAST-ADL already change the grammar significantly, causing modifications of all 91 grammar rules and changes in nearly every line of the grammar. This also illustrates how massive the changes to the generated grammar are to reach the desired grammar. The number of changes is even larger with the full version of EAST-ADL.

We only needed to change and add a total of 10 grammar optimization rule applications to complete the optimization of the grammar of full EAST-ADL. \revised{For example, we excluded the primary type \texttt{String0} from the full version of the EAST-ADL grammar, which led us to add a line of configuration \texttt{go.removeRule(String0)}}. While this is increasing the \grammaroptimizer configuration from the simple EAST-ADL version quite a bit (from 22 optimization rule applications to 31 optimization rule applications), the increase is fairly small given that the meta-model increased massively (with 200 additional metaclasses).


The reason is that our grammar optimization requirements for the simplified version and the full version of EAST-ADL are almost the same. This optimization requirement is mainly based on the look and feel of the language and is provided by an industrial partner. These optimization rule applications have been configured for the simplified version. When we applied them to the generated grammar of the full version of EAST-ADL, we found that we can reuse all of these optimization rule applications. Furthermore, we benefit from the fact that many optimization rule applications are formulated for the scope of the whole grammar and thus can also influence grammar rules added during the evolution step. We do not list a number of grammar rules in a expert-created grammar of EAST-ADL in Table~\ref{tab:result-of-supporting-evolution}, because there is no ``original'' text grammar of EAST-ADL. Instead, we optimize the generated grammar of EAST-ADL according to our industrial partner's requirements for EAST-ADL's textual concrete syntax.


\paragraph{QVTo}
The baseline configuration of the \grammaroptimizer for QVTo includes 733 optimization rule applications, which is a lot given that the expert-created grammar of QVTo 1.0 has 115 non-terminals. 
Note that the optimized grammar has even fewer grammar rules (77) as some of the rules in the optimized grammar \emph{imitate} multiple rules from the expert-created grammar at once. This again is a testament to how different the expert-created grammar is from the generated one (over 228 lines in the grammar are modified, 2 lines are added, and 580 lines are deleted by these 733 optimization rule applications).

However, if we look at the evolution towards versions 1.1, 1.2, and 1.3 we witness that very few changes to the \grammaroptimizer configuration are required. 
In fact, only between 0 and 2 out of the 733 optimization rule applications needed adjustments. 
\revised{This significantly reduces the effort required compared to manually modifying a grammar generated from a new version of the QVTo metamodel, which would require modifying hundreds of lines.}
The reason is that, even though there are many differences between different versions of the QVTo meta-model, there are only 0 to 2 differences that affect the optimization rule applications. 

For example, version 1.0 of the QVTo meta-model has an attribute called \texttt{bindParameter} in the class \texttt{VarParameter}, whereas it is called \texttt{representedParameter} in version 1.1. 
This attribute is not needed according to the expert-created grammars, so the \grammaroptimizer configuration includes a call to the optimization rule \emph{RemoveAttribute} to remove the grammar line that was generated based on that attribute.
The second parameter of the optimization rule \emph{RemoveAttribute} needs to specify the name of the attribute. As a consequence of the evolution, we had to change that name in the optimization rule application. 
Another example concerns the class \texttt{TypeDef}, which contains an attribute \texttt{typedef\_condition} in version 1.2 of the QVTo meta-model. We added square brackets to it by applying the optimization rule \emph{AddSquareBracketsToAttr} in the grammar optimization. However, in version 1.3 of the QVTo meta-model, the class \texttt{TypeDef} does not contain such an attribute, so the optimization rule application \emph{AddSquareBracketsToAttr} was unnecessary.

Most of the differences between different versions of the meta-model do not lead to changes in the optimization rule applications. For example, the multiplicity of the attribute \texttt{when} in the class \texttt{MappingOperation} is different in version 1.0 and 1.1. We used \emph{RemoveAttribute} to remove the attribute during the optimization of grammar version 1.0. The same command can still be used in version 1.1, as the removal operation does not need to consider the multiplicity of an attribute. Therefore, this difference does not affect the configuration of optimization rule applications.

\begin{table*}
\scriptsize
    \centering
    \caption{Result of supporting evolution (RQ2)}
    \label{tab:result-of-supporting-evolution}
    \begin{threeparttable}
    \begin{tabular}{@{}llrrrrrrrrrrrrrrrr@{}}
        \toprule
		~ & \textbf{Meta-m.} & \multicolumn{3}{c}{\textbf{Generated grammar}} & \multicolumn{3}{c}{\textbf{Optimized grammar}} & \multicolumn{3}{c}{\textbf{Grammar rules}} &
        \multicolumn{3}{c}{\textbf{Lines in Grammar}} \\
		\textbf{DSL} & \textbf{Classes}~\tnote{1} & \textbf{lines}	& \textbf{rules}	& \textbf{calls} & \textbf{lines} & \textbf{rules} & \textbf{calls}~\tnote{2} & \textbf{mod} & \textbf{add} & \textbf{del} & \textbf{mod} & \textbf{add} & \textbf{del} & \textbf{$\#GORA$} 	&\textbf{$\#cORA$} \\
        \midrule
        EAST-ADL & 91 & 755 & 91  & 735   & 767   & 103   & 782   & 70    & 12    & 0     & 517   & 14    & 2   & 22  & / \\
				(simple) & ~ & ~ & ~  & ~ & ~			& ~			& ~ 		& ~			& ~			&	~ 		& ~			& ~			& ~   	& ~		& ~ \\
        EAST-ADL  & 291 & 2,839 & 291 & 3,062 & 2,851 & 303   & 3,074 & 233  & 12    & 1     & 2,046 & 16    & 4   & 31  & 10\\
				(full)  & ~ & ~ & ~  & ~ & ~			& ~			& ~ 		& ~			& ~			&	~ 		& ~			& ~			& ~   	& ~		& ~ \\
        \midrule
        QVTo 1.0     & 85  & 1,026 & 109   & 910   & 444   & 77    & 181   & 66    & 1     & 33    & 228   & 2     & 580 & 733 & / \\
        QVTo 1.1     & 85  & 992   & 110   & 836   & 444   & 77    & 181   & 66    & 1     & 34    & 228   & 2     & 546 & 733 & 2 \\
        QVTo 1.2     & 85  & 992   & 110   & 836   & 444   & 77    & 181   & 66    & 1     & 34    & 228   & 2     & 546 & 733 & 0 \\
        QVTo 1.3     & 85  & 991   & 110   & 835   & 443   & 77    & 180   & 66    & 1     & 34    & 228   & 2     & 546 & 733 & 1 \\
        \bottomrule
    \end{tabular}
    \begin{tablenotes}
        \item[1] The number is after adaptation, and it contains both classes and enumerations.
        \item[2] The number includes the calls to dummy OCL and dummy SML expressions. 
    \end{tablenotes}
    \end{threeparttable}
\end{table*}

\section{Discussion}
\label{sec:discussion}
In the following, we discuss the threats to validity of the evaluation, different aspects of the \grammaroptimizer, and future work implied by the current limitations.

\subsection{Threats to Validity}\label{sec:threats}
The threats to validity structured according to the taxonomy of Runeson et al. \cite{Runeson.2008,Runeson.2012} are as follows.

\subsubsection{Construct Validity}\label{sec:threats_construct}
We limited our analysis to languages for which we could find meta-models in the Ecore format. Some of these meta-models were not ``official'', in the sense that they had been reconstructed from a language in order to include them in one of the ``zoos''. An example of that is the meta-model for BibTeX we used in our study. 
In the case of the DOT language, we reconstructed the meta-model from an Xtext grammar we found online. 
We adopted a reverse-engineering strategy where we generated the meta-model from the expert-created grammar and then generated a new grammar out of this meta-model.
This poses a threat to validity since many of the languages we looked at can be considered ``artificial'' in the sense that they were not developed based on meta-models. 
However, we do not think this affects the construct validity of our analysis since our purpose is to analyze what changes need to be made from an Xtext grammar file that has been generated.
In addition, we address this threat to validity by also including a number of languages (e.g., Xenia and Xcore) that are based on meta-models and using the meta-models provided by the developers of the language.

\revised{Furthermore, we had to make some changes to some of the meta-models to be able to generate Xtext grammars out of them at all (cf.~\Cref{sec:methodology_analysis_MMPrepsAndGrammarGen}) or to introduce certain language constructs required by the textual concrete syntax (cf.~\Cref{sec:support_eval_QVTo_Preparation}).
These meta-model adaptations might have introduced biased changes and thereby impose a threat to construct validity.
However, we reduced these adaptations to a minimum as far as possible to mitigate this threat and documented all of them in our supplemental material~\cite{datasource2023go} to ensure their reproducibility.}

\subsubsection{Internal Validity}\label{sec:threats_internal}
In the evaluation (cf.~\Cref{sec:eval}), we set up and quantitatively evaluate size and complexity metrics regarding the  considered meta-models and grammars as well as regarding the \grammaroptimizer configurations for the use cases of one-time grammar adaptations and language evolution.
Based on that, we conclude and argue in \Cref{subsubsection:eval:imtatio:results,sec:discussion_effort} about the effort required for creating and evolving languages as well as the effort to create and re-use \grammaroptimizer configurations.
These relations might be incorrect.
However, the applied metrics provide objective and obvious indications about the particular sizes and complexities and thereby the associated engineering efforts. 

\subsubsection{External Validity}\label{sec:threats_external}
As discussed in the analysis part, we analyzed a total of seven DSLs to identify generic optimization rules. Whereas we believe that we have achieved significant coverage by selecting languages from different domains and with very different grammar structures, we cannot deny that analysis of further languages could have led to more
optimization rules.
However, due to the extensible nature of \grammaroptimizer{}, the practical impact of this threat to generalisability is low since it is easy to add additional generic optimization rules once more languages are analyzed.

\revisedNew{Generalisability is further affected by the question of how representative are for other cases represented in practice. Our evaluation would be most insightful if the considered languages resemble typical practical cases, instead of corner cases. 
A nuanced assessment of how typical our considered cases are for other cases would require systematic studies of evolution histories of metamodel-driven DSLs, which, to, our knowledge, are not available yet and would be a worthwhile direction for future work.
}

\subsubsection{Reliability}\label{sec:threats_reliability}
Our overall procedure to conceive and develop the \grammaroptimizer encompassed multiple steps.
That is, we first determined the differences between the particular initially generated Xtext grammars and the grammars of the actual languages in two iterations as described in \Cref{sec:methodology}.
This analysis yielded the corresponding identified conceptual grammar optimization rules summarized in \Cref{sec:identified-optimization-rules}.
Based on these identified conceptual grammar optimization rules, we then implemented them as described in \Cref{sec:solution}.
This procedure imposes multiple threats to reliability.
For example, analyzing a different set of languages could have led to a different set of identified optimization rules, which then would have led to a different implementation.
Furthermore, analyzing the languages in a different order or as part of different iterations could have led to a different abstraction level of the rules and thereby a different number of rule.
Finally, the design decisions that we made during the identification of the conceptual optimization rules and during their implementation could also have led to different kinds of rules or of the implementation.
However, we discussed all of these aspects repeatedly amongst all authors to mitigate this threat and documented the results as part of our supplemental material~\cite{datasource2023go} to ensure their reproducibility.

\subsection{The Effort of Creating and Evolving a Language with the \grammaroptimizer }\label{sec:discussion_effort}
The results of our evaluation show three things.
%
\revised{First, the expert-created grammars of all studied languages differ greatly in appearance from the generated grammars.}
Thus, in most cases, creating a DSL with Xtext will require the language engineer to perform big changes to the generated grammar.
Second, depending on the language, using the \grammaroptimizer for a single version of the language may or may not be more effort for the language engineer, compared to manually adapting the grammar.
Third, there seems to be a large potential for the reuse of \grammaroptimizer configurations between different versions of a language, thus supporting the evolution of textual languages.

These observations can be combined with the experience that most languages evolve with time and that especially DSLs go through a rapid prototyping phase at the beginning where language versions are built for practical evaluation~\cite{wang2005rapidly}. 
Therefore, we conclude that the \grammaroptimizer has big potential to save manual effort when it comes to developing DSLs.

\revised{Additionally, a topic worth mentioning is how the involvement of different people and their skill sets affect the effort when creating and reusing optimization rule configurations. For example, in case that updates to an existing configuration are needed after an evolution step, the maintainers need to understand the optimization rule configuration of the previous version, which could take a new contributor more time than the original contributor. Assessing the impact of this aspect is a subject for  future work.}

\subsection{Implications for Practitioners and Researchers}
Our results have several implications for language engineers and researchers. 

\paragraph{Impact on Textual Language Engineering}	
Our work might have an impact on the way DSL engineers create textual DSLs nowadays. 
That is, instead of specifying grammars and thereby having to be EBNF experts, the \grammaroptimizer also enables engineers familiar with meta-modelling to conceive well-engineered meta-models and to semi-automatically generate user-friendly grammars from them.
Furthermore, Kleppe \cite{Kleppe.2007b} compiles a list of advantages of approaches like the \grammaroptimizer, among them two that apply especially to our solution: 1) the \grammaroptimizer provides flexibility for the DSL engineering process, as it is no longer necessary to define the kind of notation used for the DSL at the very beginning
as well as 2) the \grammaroptimizer enables rapid prototyping of textual DSLs based on meta-models.

\paragraph{Blended Modeling}
Ciccozzi et al.~\cite{Ciccozzi.2019} coin the term \emph{blended modeling} for the activity of interacting with one model through multiple notations (e.g., both textual and graphical notations), which would increase the usability and flexibility for different kinds of model stakeholders. 
However, enabling blended modeling shifts more effort to language engineers.
This is due to the fact that the realization of the different editors for the different notations requires many manual steps when using conventional modeling frameworks.	
In this context, Cicozzi and colleagues particularly stress the issue of the manual customization of grammars in the case of meta-model evolution.
Thus, as one research direction to enable blended modeling, Ciccozzi et al.\ formulate the need to automatically generate the different editors from a given meta-model.	
Our work serves as one building block toward realizing this research direction and opens up the possibility to develop and evolve blended modeling languages that include textual versions.

\revised{A relevant question is to which extent our approach enables cost savings in a  larger context, as the   cost for evolving the existing tools and applications working with existing languages might be higher than the cost for evolving the languages themselves.
We benefit from the extensive tool support offered by Xtext, which can automatically re-generate large parts of the available textual editor after changes of the underlying grammar, including features such as, e.g., auto-formatting, auto-completion, and syntax highlighting. In consequence, by supporting automated grammar changes (in particular, after evolution steps), we also save effort for the overall adaptation of the textual editor. However, in MDE contexts, other applications and tools typically refer to the metamodel, instead of the grammar, and hence, are outside our scope.}

\paragraph{Prevention of Language Flaws}
Willink~\cite{willink2020reflections} reflects on the version history of the Object Constraint Language (OCL) and the flaws that were introduced during the development of the different OCL 2.x specifications by the Object Management Group~\cite{OCLVersions}.
Particularly, he points out that the lack of a parser for the proposed grammar led to several grammar inaccuracies and thereby to ambiguities in the concrete textual syntax.
This in turn led to the fact that the concrete syntax and the abstract syntax in the Eclipse OCL implementation \cite{EclipseOCL} are so divergent that two distinct meta-models with a dedicated transformation between both are required, which also holds for the QVTo specification and its Eclipse implementation \cite{willink2020reflections} (cf.~\Cref{sec:background}).
The \grammaroptimizer will help to prevent and bridge such flaws in language engineering in the future. 
Xtext already enables the generation of the complete infrastructure for a textual concrete syntax from an abstract syntax represented by a meta-model. Our approach adds the ability to optimize the grammar (i.e., the concrete syntax), as we show in the evaluation by deriving an applicable parser with an optimized grammar from the QVTo specification meta-models.	

\subsection{Future Work}
The \grammaroptimizer is a first step in the direction of supporting the evolution of textual grammars for DSLs. However, there are, of course, still open questions and challenges that we discuss in the following.

\paragraph{Name Changes to Meta-model Elements}
In the \grammaroptimizer configurations, we currently reference the grammar concepts derived from the meta-model classes and attributes by means of the class and attribute names (cf.~\Cref{lst:dot-config-example-rule}).
Thus, if a meta-model evolution involves many name changes, likewise many changes to optimization rule applications are required. 
Consequently, we plan as future work to improve the \grammaroptimizer with a more flexible concept, in which we more closely align the grammar optimization rule applications with the meta-model based on name-independent references.

\paragraph{More Efficient Rules and Libraries}
We think that there is a lot of potential to make the available set of optimization rules more efficient. 
This could for example be done by providing libraries of more complex, recurring changes that can be reused. 
\revised{Such a library can contain a default set of optimization rule configurations to make the generated grammar follow a particular style (e.g., mimicking an existing language, to be appealing for users of that language). Language engineers can use it as a basis and with minimal effort define optimization rule configurations that perform DSL-specific changes.}
Such a change might make the application of the \grammaroptimizer attractive even in those cases where no evolution of the language is expected.
\revised{While this use-case still requires effort for defining configurations, the overall effort compared to manual editing can be reduced especially in cases with applicable large-scoped rules that, e.g., globally change the parenthesis style in the grammar. }

In addition, the API of \grammaroptimizer could be changed to a fluent version where the optimization rule application is configured via method calls before they are executed instead of using the current API that contains many \texttt{null} parameters.
This could also lead to a reduction of the number of grammar optimization rule applications that need to be executed since some executions could be performed at the same time.

Another interesting idea would be to use artificial intelligence to learn existing examples of grammar optimizations in existing languages to provide optimization suggestions for new languages and even automatically create configurations for the \grammaroptimizer. 


\paragraph{Expression Languages}
In this paper, we excluded the expression language parts (e.g., OCL) of two of the example languages (cf.~\Cref{subsubsect:Meth:exlusionParts}). 
However, expression languages define low-level concepts and have different kinds of grammars and underlying meta-models than conventional languages. 
In future work, we want to further explore expression languages specifically, in order to ensure that the \grammaroptimizer can be used for these types of syntaxes as well.

\paragraph{Visualization of Configuration}
Currently, we configure the \grammaroptimizer by calling the methods of optimization rules, which is a code-based way of working. In the future, 
we intend to improve the tooling for \grammaroptimizer and embed the current library into a more sophisticated workbench that allows the language engineer to select and parameterize optimization rule applications either using a DSL or a graphical user interface and provides previews of the modified grammar as well as a view of what valid instances of the language look like.

\paragraph{Co-evolving Model Instances}\label{sec:discussion_coevolution_instances}
We also intend to couple \grammaroptimizer with an approach for language evolution that also addresses the model instances. 
\revised{In principle, a model instance represented by a textual grammar instance can be} read using the old grammar and parsed into an instance of the old meta-model. It can then be transformed, e.g., using QVTo to conform to the new meta-model, and then be serialized again using the new grammar. However, following this approach means that formatting and comments can be lost. Instead, we intend to derive a textual transformation from the differences in the grammars and the optimization rule applications that can be applied to the model instances and maintain formatting and comments as much as possible.

\revised{\paragraph{Alternative implementation strategy}\label{sec:discussion_alternative_impl}
Our implementation strategy relies on the format of textual grammars produced by Xtext, which is  stable across recent versions of Xtext.
This implementation strategy was suitable for positively answering our evaluation questions and thus, substantiating the scientific contribution of our paper.
An alternative, arguably more elegant implementation strategy would be to use Xtext's abstract syntax tree representation of the grammar.
A benefit of such an implementation would be that it would be more robust in case that the output format of Xtext changes, rendering it a desirable direction for future work.}




\section{Conclusion}
In this paper, we have presented \grammaroptimizer, a tool that supports language engineers in the rapid prototyping and evolution of textual domain-specific languages which are based on meta-models. \grammaroptimizer uses a number of optimization rules to modify a grammar generated by Xtext from a meta-model. These optimization rules have been derived from an analysis of the difference between the actual and the generated grammars of seven DSLs.

We have shown how \grammaroptimizer can be used to modify grammars generated by Xtext based on these optimization rules. This automation is particularly useful while a language is being developed to allow for rapid prototyping without cumbersome manual configuration of grammars and when the language evolves. We have evaluated \grammaroptimizer on seven grammars to gauge the feasibility and effort required for defining the optimization rules. We have also shown how \grammaroptimizer supports evolution with the examples of EAST-ADL and QVTo.

Overall, our tool enables language engineers to use a meta-model-based language engineering workflow and still produce high-quality grammars that are very close in quality to hand-crafted ones. We believe that this will reduce the development time and effort for domain-specific languages and will allow language engineers and users to leverage the advantages of using meta-models, e.g., in terms of modifiability and documentation.

In future work, we plan to extend \grammaroptimizer into a more full-fledged language workbench that supports advanced features like refactoring of meta-models, a ``what you see is what you get'' view of the optimization of the grammar, and the ability to co-evolve model instances alongside the underlying language. We will also explore the integration into workflows that generate graphical editors to enable blended modelling.

\section*{Acknowledgements}
This work has been sponsored by Vinnova under grant number 2019-02382 as part of the ITEA 4 project \emph{BUMBLE}.


\bibliographystyle{elsarticle-num-names}
\bibliography{GrammarOptimizer}

\end{document}